\documentclass[twoside]{IEEEtran}

\usepackage{graphicx}
\usepackage[tight,footnotesize]{subfigure}
\usepackage{array}
\usepackage[cmex10]{amsmath}
\usepackage{amssymb,amsthm,amsfonts}
\usepackage{array}
\usepackage{cite}
\usepackage{color}

\newtheorem{theorem}{Theorem}
\newtheorem{lemma}[theorem]{Lemma}
\theoremstyle{remark}
\newtheorem{remark}{Remark}

\newcommand{\vecr}[1]{\textbf{\uppercase{#1}}}
\newcommand{\vecd}[1]{\textbf{\lowercase{#1}}}
\newcommand{\matd}[1]{{\sf{#1}}}
\newcommand{\matr}[1]{{\mathbb{#1}}}
\newcommand{\const}[1]{{\mathcal{#1}}}
\newcommand{\Reals}{\mathbb{R}}
\newcommand{\Complex}{\mathbb{C}}

\newcommand{\normalr}[2]{\mathcal{N}_{\Reals}\!\left(#1,#2\right)}
\newcommand{\E}{\mathsf{E}}
\newcommand{\thmref}[1]{Theorem~\ref{#1}}

\newcommand{\remref}[1]{Remark~\ref{#1}}

\newcommand{\SNR}{\textnormal{SNR}}

\newcommand{\set}[1]{\mathcal{#1}}

\begin{document}

\title{On the Per-Sample Capacity of\\Nondispersive Optical Fibers}
\author{Mansoor~I.~Yousefi and~Frank~R.
Kschischang,\IEEEmembership{~Fellow,~IEEE}
\thanks{This paper was presented in part
at the 25th Biennial Symposium on Communications, Kingston, ON, in May 2010,
at the 2010 IEEE International Symposium on Information Theory,
Austin, TX, in June 2010, and
at the 2011 Canadian Workshop on Information Theory, Kelowna, BC, in May 2011.}
\thanks{The authors are with the Edward~S.~Rogers~Sr.\ Department of
Electrical and Computer Engineering, University of Toronto, Toronto,
ON M5S 3G4, Canada, e-mails: {\tt \{mansoor,frank\}@comm.utoronto.ca}.}
}
\date{}
\markboth{IEEE Transactions on Information Theory}{Yousefi and Kschischang:
Per-Sample Capacity of Nondispersive Optical Fibers}
\IEEEpubid{0000--0000/00\$00.00~\copyright~2011 IEEE}

\maketitle

\begin{abstract}
The capacity of the channel defined by the stochastic nonlinear
Schr\"odinger equation, which includes the effects of the Kerr
nonlinearity and amplified spontaneous emission noise, is considered
in the case of zero dispersion.  In the absence of dispersion, this
channel behaves as a collection of parallel per-sample channels.  The
conditional probability density function of the nonlinear per-sample
channels is derived using both a sum-product and a Fokker-Planck
differential equation approach. It is shown that, for a fixed noise
power, the per-sample capacity grows unboundedly with input signal.
The channel can be partitioned into amplitude and phase subchannels,
and it is shown that the contribution to the total capacity of the
phase channel declines for large input powers. It is found that a
two-dimensional distribution with a half-Gaussian profile on the
amplitude and uniform phase provides a lower bound for the
zero-dispersion optical fiber channel, which is simple and
asymptotically capacity-achieving at high
signal-to-noise ratios (SNRs).
A lower bound on the
capacity is also derived in the medium-SNR region. The exact capacity
subject to peak and average power constraints is numerically
quantified using dense multiple ring modulation formats. The
differential model underlying the zero-dispersion channel is reduced
to an algebraic model, which is more tractable for digital
communication studies, and in particular it provides a relation
between the zero-dispersion optical channel and a $2 \times 2$
multiple-input multiple-output Rician fading channel. It appears that
the structure of the capacity-achieving input distribution resembles
that of the Rician fading channel, \emph{i.e.}, it is discrete in
amplitude with a finite number of mass points, while continuous and
uniform in phase.
\end{abstract}

\begin{IEEEkeywords}
Information theory, optical fiber, nonlinear Schr\"odinger equation,
path integral.
\end{IEEEkeywords}

\maketitle

\section{Introduction}

\IEEEPARstart{A}{lthough}
the capacity of many classical communication channels has
been established, determining the capacity of fiber-optic channels has
remained an open and challenging problem.  The capacity of the optical
fiber channel is difficult to evaluate because signal propagation in
optical fibers is governed by the stochastic nonlinear Schr{\"o}dinger
(NLS) equation, which causes signal and noise to interact in a
complicated way.  This paper evaluates the capacity for models of the
optical fiber channel in the case of zero dispersion.

The deterministic NLS equation is a partial differential equation in
space and time exhibiting linear dispersion and a cubic nonlinearity,
giving rise to a deterministic model of pulse propagation in optical
fibers in the absence of noise.  When distributed additive white noise
is incorporated, a waveform communication channel is defined.  This
channel has an input-output map that is not explicit and
instantaneous, but involves the evolution of the transmitted signal
along the space dimension.

The exact capacity of the optical fiber with dispersion and
nonlinearity is not yet known. Results so far are limited to lower
bounds on the capacity in certain regimes of propagation or under some
conditions.  Some of these works include modeling the nonlinearity by
multiplicative noise in the wavelength-division multiplexing (WDM)
case \cite{mitra2000nli}, assuming a Gaussian distribution for the
output signal when nonlinearity is weak \cite{tang2001scc}, perturbing
the nonlinearity parameter \cite{narimanov2002ccf}, approaching
capacity via multiple-ring modulation formats \cite{essiambre2010clo},
or specializing to the important case of zero dispersion
\cite{tang2001scc,turitsyn2003ico}, which is the focus of this paper.

When dispersion is zero, pulse propagation is governed only by the
Kerr nonlinearity and amplified spontaneous emission (ASE) noise.
This eliminates the time-dependence of the stochastic NLS, reducing it
to a nonlinear ordinary differential equation (ODE) as a function only
of distance $z$. For a certain suboptimal receiver, as assumed later
in the paper, transmission is then sample-wise and the channel can be
viewed as a collection of parallel independent sub-channels, with
noise interacting with the nonlinearity in the same channel, but not
with neighboring channels located at other times.  The problem becomes
easier to analyze since, instead of describing the evolution of a
random waveform and its entropy rate, we merely need to look at the
evolution of a random variable and its entropy, which can be described
by a single conditional probability density function (PDF).

\IEEEpubidadjcol

In \cite{tang2001scc}, Tang estimated the capacity in the
dispersion-free case using Pinsker's formula, based on the channel
input-output correlation functions.  Tang's results show that
capacity, $\const{C}$, increases with input power power, $\const{P}$,
reaching a peak at a certain optimal input power, and then
asymptotically vanishes as $\const{P} \rightarrow \infty$.  Estimates
of the capacity in the general case
\cite{mitra2000nli,narimanov2002ccf,kahn2004sel,essiambre2010clo} also
exhibit this behavior. However, the results of \cite{tang2001scc} can
be viewed only as a lower bound on the capacity (even when the
nonlinearity is weak), since the second-order statistics used in
Pinsker's formula do not capture the entire conditional PDF, which, of
course, is required for the computation of $\const{C}$.

The conditional probability density function (of the channel output
given the channel input) for the dispersion-free optical fiber has
been derived in \cite{turitsyn2003ico} and \cite{mecozzi1994llh}. In
\cite{turitsyn2003ico}, the authors used the Martin-Siggia-Rose
formalism in quantum mechanics to find a closed-form expression for
the conditional joint PDF of the received signal amplitude and phase.
Although they did not explicitly compute the capacity, they showed
that the capacity asymptotically goes to infinity for large
signal-to-noise ratios (\SNR s). 

With the exception of a few papers (\emph{e.g.},
\cite{mitra2000nli,tang2001scc,narimanov2002ccf,turitsyn2003ico,kahn2004sel,essiambre2010clo}),
optical fiber communication is largely unstudied from the information
theory point of view.  Most previous papers on the capacity of optical
fibers have focused on the dispersive channel directly from its
description given by the stochastic nonlinear Schr\"odinger equation.
This direct approach has had limited success, due to the complexity of
the underlying channel model and its limited mathematical
understanding. The stochastic nonlinear Schr{\"o}dinger equation with
dispersion parameter set to zero, on the other hand,  leads to a model
which is the basic building block of the dispersive optical fiber
channel. It is therefore of fundamental interest to first study the
zero-dispersion case.  In this paper, we pursue such a bottom-up
approach. Below, we highlight some of the contributions of this paper.

In Sec.~\ref{subsec:recursive}, we provide a simple derivation of the
conditional PDF of the channel output given the channel input.  Our
approach is based on discretizing the fiber into a cascade of a large
number of small fiber segments, which leads to a recursive computation
of the PDF.  An alternative perspective, using a stochastic calculus
approach, is provided in Appendix~\ref{sec:fokkerplanck}.

In Sec.~\ref{subsec:sumproduct}, we show that the probabilistic
channel model in optical fibers can be understood in terms of the
sum-product algorithm, or as a path integration.  Such path integrals
underlie the Martin-Siggia-Rose formalism, which was employed in
\cite{turitsyn2003ico}.

In Sec.~\ref{sec:num-eval}, for the first time to the best of our
knowledge, the capacity of the dispersionless fiber is numerically
evaluated as a function of the signal-to-noise ratio, for fixed noise
spectral densities.  The results re-affirm the conclusion of
\cite{turitsyn2003ico} that the channel capacity (when measured in
bits per symbol) grows unbounded, at a fixed noise level, with
increasing signal power.

In Sec.~\ref{sec:simmod}, a decomposition is established between the
amplitude and phase channels.  The decoupling has the property that
the input phase does not statistically excite the output amplitude.
Using this amplitude and phase decomposition,
the asymptotic result of \cite{turitsyn2003ico} is then easily proved.

Also in Sec.~\ref{sec:simmod}, a simplified model is derived for the
dispersion-free optical channel.  Simplification is achieved by
reducing the \emph{differential model} underlying the zero-dispersion
channel to an \emph{algebraic model}, which is more tractable for
digital communication studies. Instead of a stochastic differential
equation, the channel's input/output relation is explicitly expressed
as a simple 2$\times$2 MIMO system, similar to MIMO wireless
multi-antenna models.

In Sec.~\ref{sec:fundlimits} we return to the amplitude/phase
decomposition, and show that the phase channel exhibits a \emph{phase
transition property}: for very small or high signal power levels, the
phase channel conveys little information, and the maximum information
rate is achieved at a finite optimal power. This phase transition
property partitions the $\const{P}$-$\sigma^{2}$ plane into four
regions which behave in different ways. This partitioning also enables
us to find practically significant bounds on the capacity in some of
these regions. 

In \cite{mecozzi2001cim} it was shown that for a simple
intensity-modulation direct-detection (IM/DD) optical channel, a
half-Gaussian distribution is asymptotically capacity-achieving at
high SNRs. An important conclusion of Sec.~\ref{sec:fundlimits} is
that, a two-dimensional distribution with a half-Gaussian profile on
the amplitude and uniform phase provides an excellent global lower
bound for the zero-dispersion optical fiber channel, which is simple
and asymptotically capacity-achieving in a certain high SNR regime
where $\const{P}\rightarrow\infty$ and noise power is fixed. 

In Sec.~\ref{sec:two-d}, we show that the channel capacity is indeed a
two-dimensional function of the signal and noise powers and, unlike
classical linear channels, is not completely captured by the
signal-to-noise ratio.

In Sec.~\ref{sec:speceff}, we address the relationship between the
spectral efficiency in bits/s/Hz and capacity in bits/symbol.  For
many practical pulse shapes, even though the capacity (in bits/symbol)
grows without bound (in agreement with \cite{turitsyn2003ico}),
spectral broadening resulting from the fiber nonlinearity sends the
spectral efficiency (in bits/s/Hz) to zero (in agreement with the
estimates of \cite{tang2001scc}).  This result also agrees with the
spectral-efficiency estimates of
\cite{mitra2000nli,narimanov2002ccf,kahn2004sel,essiambre2010clo} for
fiber channels with nonzero dispersion.

In Sec.~\ref{sec:inputdist}, the optical fiber at zero dispersion is
related to the Rician fading channel in wireless communication.
Although we do not provide a formal proof, numerical simulations
indicate that the optimal capacity-achieving input distribution for
the dispersion-free optical fiber appears to be discrete in amplitude
and uniform in phase.

\section{Notation}

The notation in this paper is mostly consistent with \cite{moser2004dbb}.  We
use upper-case letters to denote scalar random variables taking values
on the real line $\mathbb{R}$ or in the complex plane $\mathbb{C}$,
and lower-case letters for their realizations. Random vectors are
denoted with
bold-face capital letters while their realizations are
denoted by boldface lower-case letters. All deterministic quantities
are treated as realizations of random variables. In order to avoid
confusion with scalar random variables, we represent constant matrices
with sans serif font, such as $\sf{K}$, $\sf{M}$, $\sf{P}$, and
important scalars with calligraphic font  such as power $\mathcal{P}$,
bandwidth $\mathcal{W}$, capacity $\mathcal{C}$, rate $\mathcal{R}$.
We reserve lower case Greek and Roman letters for special scalars.
Real and complex normal random variables are shown as
$\mathcal{N}_{\mathbb{R}}$ and $\mathcal{N}_{\mathbb{C}}$. We use the
shorthand notation $\{\textbf{Z}_{k}\}\sim\textrm{IID}$
$\mathcal{N}_{\mathbb{R}}(0,\sigma^{2})$ to denote a sequence of real,
independent, identically-distributed zero-mean Gaussian random
variables with variance $\sigma^{2}$.

\section{Channel Model} \label{sec:channelmodel}

Let $Q(z,t)$ be the complex envelope of the propagating electric field
as a function of distance $z$ and time $t$, the latter measured with
respect to a reference frame copropagating with the signal.  Signal
evolution in optical fibers with zero dispersion and distributed Raman
amplification is modeled by the stochastic nonlinear ODE (see,
\emph{e.g.}, \cite[Eqn.~(1) with $\beta_2=0$]{essiambre2010clo})
\begin{IEEEeqnarray}{rCl}
  \frac{\partial Q(z,t)}{\partial z}
  & = & j\gamma Q(z,t)|Q(z,t)|^{2}+V(z,t), \nonumber \\
  Q(0,t) & = & Q_{0}(t),\quad 0 \leq z \leq \const{L}.
  \label{eq:zerodisw}
\end{IEEEeqnarray}
Here $\const{L}$ is the length of the fiber, $V(z,t)$ is a zero-mean
Gaussian process uncorrelated in space and time, \emph{i.e.}, with
\[
\E[V(z,t)V^*(z^{\prime},t^\prime)]=\sigma_0^{2}\delta(z-z^{\prime},t-t^\prime),
\]
and $Q_0(t)$ is the complex envelope of the electric field applied at
the fiber input. Finally, $\gamma$ is the Kerr nonlinearity parameter
and $\sigma_0^2$ is the noise power spectral density.  Following
\cite{essiambre2010clo}, we assume the fiber parameters given in
Table~\ref{tbl:fiberparam}, with $\sigma_0^2 = n_{\rm sp} h \nu \alpha
= 5.906 \times 10^{-21}~{\rm W/(km}\cdot{\rm Hz)}$. The transmitted
power is denoted as
\[
P_{Q_0} =
\frac{1}{\const{T}}\int_{0}^{\const{T}}|Q_{0}(t)|^{2}dt.
\]
The transmitter will be constrained so that $\E P_{Q_0} =
\const{P}_{0}$.

The stochastic differential equation (\ref{eq:zerodisw}) is
interpreted in the It\^{o} sense via its equivalent integral
representation \cite{gardiner1985hsm}.  Note that in
(\ref{eq:zerodisw}) the white noise $V(z,t)$ is added to the spatial
derivative of the signal, as opposed to the signal itself, and has
units of ${\rm W}^{1/2}/{\rm km}$.  Note further that
(\ref{eq:zerodisw}) contains no loss parameter, since losses are
assumed to be perfectly compensated for by Raman amplification
\cite{islam2003rat}.  The time variable $t$ appears in
(\ref{eq:zerodisw}) essentially as a parameter. Some limitations of
the zero-dispersion model \eqref{eq:zerodisw} are discussed in
Section~\ref{sec:speceff}.

\begin{table}[b]
\caption{Fiber Parameters}
\label{tbl:fiberparam}
\centerline{\begin{tabular}{c|l|l}
$n_{\rm sp}$ & 1 & {\footnotesize spontaneous emission factor}\\
$h$ & $6.626 \times 10^{-34} {\rm J} \cdot {\rm s}$ & {\footnotesize Planck's constant} \\
$\nu$ & 193.55~{\rm THz} & {\footnotesize center frequency} \\
$\alpha$ & $0.046~{\rm km}^{-1}$ & {\footnotesize fiber loss (0.2~dB/km)} \\
$\gamma$ & $1.27~{\rm W}^{-1}{\rm km}^{-1}$ & {\footnotesize nonlinearity parameter} \\
$B$ & 125~GHz & {\footnotesize maximum bandwidth} \\
\end{tabular}}
\end{table}

Suppose the communication channel (\ref{eq:zerodisw}), \emph{i.e.},
the waveform channel from $Q(0,t)$ to $Q(\const{L},t)$, is used in the
time interval $[0,\const{T}]$ for some fixed $\const{T}$ and that the
input waveform $Q_0(t)$ is approximately bandlimited to $\const{W}_0$
Hz. It is well known that, if $\const{W}_{0} \const{T} \gg 1$, the set
of possible transmitted pulses spans a complex finite-dimensional
signal space (called here the input space) with approximately $2
\const{W}_0 \const{T}$ dimensions (complex degrees of freedom). Since
the channel is dispersionless, the pulse duration remains constant
during propagation; however, as discussed in Sec.~\ref{sec:speceff},
the pulse bandwidth may continuously grow because of the nonlinearity.
We denote by $\const{W}_\const{L}$ the bandwidth of the waveform
received at the output of the fiber. The received waveform is an
element of a signal space (called here the output space) of dimension
$2\const{W}_\const{L}\const{T} \ge 2\const{W}_0 \const{T}$. In other
words, the dimension of the signal space grows while the signal is
propagated.

We consider a model in which noise throughout the fiber is bandlimited
to $\const{W}_{\const{L}}$ using in-line channel filters.  Therefore,
throughout the fiber, the noise lies in the output space.  Both the
input waveform and the output waveform can be represented by samples
taken $1/2\const{W}_{\const{L}}$ seconds apart.  At the fiber input,
where the signal is constrained to lie in the input space, information
can be encoded in samples corresponding to the input signal degrees of
freedom (called the principal samples).  All other samples are
interpolated as appropriate linear combinations of the principal
samples.  These samples, though not innovative, carry correlation
information, much like parity-checks in a linear code, that are
potentially useful for optimal detection.

In this paper, however, we consider a suboptimal receiver that ignores
the additional samples, and bases its decision only on the principal
samples.  The resulting channel is consequently a set of parallel
independent scalar channels (called per-sample channels) defined via
\begin{IEEEeqnarray}{rCl}
\frac{\partial Q(z)}{\partial z}
& = & j\gamma Q(z)|Q(z)|^{2}+V(z), \nonumber \\
\E[V(z)V^{*}(z^{\prime})]&=&\sigma^{2}\delta(z-z^{\prime}), \qquad \E |Q_{0}|^{2}\leq\const{P}, \label{eq:zerodis}
\end{IEEEeqnarray}
where
$\sigma^{2}=2\const{W}_{\const{L}}\sigma_{0}^{2}$,
$Q_0 \in \mathbb{C}$ is the channel input sample value,
and $\const{P}$ is the per-sample power. The output of the channel is $Q(\const{L})\in\Complex$.

\subsection{A Simple Recursive Derivation of the Conditional PDF}
\label{subsec:recursive}

The differential model in the form given by \eqref{eq:zerodis} is not
directly suitable for an information-theoretic analysis. Instead, we
require an explicit input-output probabilistic model, \emph{i.e.}, the
conditional probability density function of the channel output given
the channel input. The conditional PDF of $Q(\const{L})$ given $Q(0)$
was derived in \cite{mecozzi1994llh} and \cite{turitsyn2003ico}.
Although the direct calculation of moments of the received signal, or
equivalently the moment generating function as in
\cite{mecozzi1994llh}, leads to an expression for the PDF, it does not
provide enough insight into the statistical nature of the channel. The
approach of \cite{turitsyn2003ico} relies on the Martin-Siggia-Rose
formalism in quantum mechanics and expresses the PDF as a path
integral. Below we derive the PDF in a simple way, by breaking the
fiber into a cascade of a large number of small segments, and
recursively compute the PDF. With this approach, we
are able to illustrate some important properties of the statistical
channel model.

In order to describe the statistics of the per-sample channels in
(\ref{eq:zerodis}), we look at the fiber as a cascade of a large
number $n \rightarrow \infty$ of pieces of discrete fiber segments by
discretizing the equation (\ref{eq:zerodis}). The recursive stochastic
difference equation giving the input-output relation of the
$k^{\textrm{th}}$ incremental channel is given by
\begin{equation} \label{dis-zerodis}
 Q_{k+1}=Q_{k}+j\epsilon\gamma|Q_{k}|^{2}+\sqrt\epsilon V_{k},\quad{0 \leq k \leq
 n-1},
\end{equation}
in which $\epsilon = \const{L}/n$, and the discrete noise $V_{i}
\sim \mathcal{N}_{\mathbb{C}}(0,\sigma^{2})$ has been scaled by the
square root of the step size, \emph{i.e.}, multiplied by $1/\sqrt
\epsilon$. Note that from (\ref{dis-zerodis}) the cascade of
incremental fiber segments forms a discrete-time continuous-state
Markov chain
\begin{equation} \label{markov}
Q_{0} \rightarrow Q_{1}\rightarrow \cdots \rightarrow Q_{n-1}.
\end{equation}

Given $Q_0$, the signal entropy is increased by a tiny,
signal-dependent, amount in each of these incremental channels. In
fact, the conditional entropy increases more for transmitted signals
with higher amplitude than those with smaller amplitude. In a
sphere-packing picture, ``noise balls'' surrounding a transmitted
symbol increase in volume as the symbol amplitude increases, and
indeed are not perfectly spherical.

Each of the incremental channels, though still nonlinear with respect
to the input signal, is conditionally Gaussian with PDF
\begin{IEEEeqnarray}{rCl} 
\label{incdens}
f_{Q_{k+1}|Q_{k}}(q_{k+1}|q_{k})&=& \frac{1}{\pi \sigma^{2}\epsilon}
\nonumber \\&& \exp\left(-\frac{\left|q_{k+1}-q_{k}-j\epsilon\gamma
q_{k}|q_{k}|^{2}\right|^2}{\epsilon
 \sigma^{2}}\right). \IEEEeqnarraynumspace
\end{IEEEeqnarray}

Using the Markov property (\ref{markov}), the probability density
function for the cascade of two consecutive incremental channels is
given by the Chapman-Kolmogorov equation
\begin{IEEEeqnarray}{rCl}
&&f_{Q_{k+2}|Q_{k}}(q_{k+2}|q_{k})= \nonumber \\
&&\int\limits_{\mathbb{C}} f_{Q_{k+2}|Q_{k+1}}(q_{k+2}|q_{k+1})
f_{Q_{k+1}|Q_{k}}(q_{k+1}|q_{k}) dq_{k+1}. \label{chap-kol}
\end{IEEEeqnarray}
Repeated application of (\ref{chap-kol}) gives the overall conditional
PDF
\begin{IEEEeqnarray}{ll}
&f_{Q_n|Q_{0}}(q_n|q_{0})=\int\limits_{\mathbb{C}} \ldots
\int\limits_{ \mathbb{C}} \frac{1}{(
 \pi \sigma^{2}\epsilon)^{n}}\nonumber \\
&{}\prod_{k=0}^{n-1}{}\:\exp\left(-\frac{\left|q_{k+1}-q_{k}-j\epsilon\gamma|q_{k}|^{2}q_{k}\right|^{2}}{\epsilon
\sigma^{2}}\right)\prod_{k=1}^{n-1}dq_{k}\label{eq:pathint1-1}\\
=&\frac{1}{\pi\sigma^2\epsilon}\int\limits_{\mathbb{C}} \ldots \int\limits_{ \mathbb{C}}
\exp\left\{-\frac{\epsilon}{\sigma^{2}}\left[\sum\limits_{k=0}^{n-1}\left|\frac{q_{k+1}-q_{k}}{\epsilon}-j\gamma|q_{k}|^{2}q_{k}\right|^{2}
 \right]\right\}\nonumber\\
 &\times\frac{dq_{n-1}}{\pi\epsilon\sigma^{2}}\frac{dq_{n-2}}{\pi\epsilon\sigma^{2}}\cdots\frac{dq_{1}}{\pi\epsilon\sigma^{2}},\label{eq:pathint1-2}
\end{IEEEeqnarray}
where $q_k=r_k\exp(j\phi_k)$ and integrations are performed over the entire complex plane.

We proceed to solve multiple integrals \eqref{eq:pathint1-2}.  First a
change of variables is introduced to make the exponent quadratic in
$q_k$. The \emph{integrating factor} for the noiseless equation
\eqref{eq:zerodis} serves as the new variable
\begin{IEEEeqnarray}{rCl}\label{eq:changvar}
p_k=q_k\exp \left(-j\epsilon\gamma
  \sum\limits_{m=0}^{k-1}\left|q_m\right|^2\right)\quad k=1,\ldots,n-1.
\end{IEEEeqnarray}

Plugging \eqref{eq:changvar} into \eqref{eq:pathint1-2}, each term in
the exponent, except the first and last terms, is
\begin{multline*}
\epsilon
\left|\frac{q_{k+1}-q_k}{\epsilon}-j\gamma|q_k|^2q_k\right|^2 \nonumber \\
=\epsilon \biggl| \frac{p_{k+1}\exp(
  j\epsilon\gamma|p_k|^2)-p_k}{\epsilon}-j\gamma|p_k|^2p_k\biggr|^2\nonumber\\
=\epsilon\left|\frac{p_{k+1}-p_k}{\epsilon}\right|^2+O(\epsilon^2),
\end{multline*}
where we have assumed $(p_{k+1}-p_k)/\epsilon$ is bounded. It follows
that each middle term in the exponent of \eqref{eq:pathint1-1} is
simplified to $\epsilon\left|\frac{p_{k+1}-p_k}{\epsilon}\right|^2$,
which is correct up
to first order in $\epsilon$.

The treatment of the boundary terms is different and in particular
they lead to new expressions. The first term is treated as above to
give
\begin{IEEEeqnarray*}{rCl}
\epsilon\biggl|\frac{p_1\exp(
  j\epsilon\gamma|q_0|^2)-q_0}{\epsilon}-j\gamma|q_0|^2q_0 \biggr|^2=\epsilon\left|\frac{p_1-q_0}{\epsilon}\right|^2+O(\epsilon^2),
\end{IEEEeqnarray*}
while the last term reads
\begin{IEEEeqnarray*}{rCl}
\epsilon \biggl|
\frac{q_n\exp\left(
  -j\gamma\epsilon\sum\limits_{k=0}^{n-1}|p_k|^2\right)-p_{n-1}}{\epsilon}\biggr|
^2+Q(\epsilon^2).
\end{IEEEeqnarray*}

As can be seen, the end point is now interpreted as
$q_n\exp(-j\epsilon\gamma\sum\limits_{k=0}^{n-1}|p_k|^2)$ which has a
variable phase. In order to consistently convert all correlated
integrals to Gaussian type integrals, one can assume that the received
phase
\begin{IEEEeqnarray}{rCl}\label{eq:phase-const}
\theta_{n}=\phi_n-\epsilon\gamma\sum\limits_{k=0}^{n-1}|p_{k}|^2,
\end{IEEEeqnarray}
is constant, and then integrate over $0\leq \theta_n<2\pi$. For fixed
$\theta_n$, the resulting integrand is an exponential with a complex
quadratic polynomial in the exponent, and boundary terms $r_0\exp
j\phi_0$ and $r_n\exp j\theta_{n}$. However phase constraint
\eqref{eq:phase-const} implies that for each value of $\theta_n$,
integration is performed under the constraint
$\epsilon\gamma\sum\limits_{k=0}^{n-1}|p_{k}|^2=\phi_n-\theta_n$.

To enforce constraint \eqref{eq:phase-const}, the function under the
integration is multiplied by the \emph{delta function} representing
that constraint, giving
\begin{IEEEeqnarray*}{rCl}
&&\delta
\left(\phi_n-\theta_{n}-\epsilon\gamma\sum\limits_{k=0}^{n-1}|p_k|^2\right)\nonumber
\\
&=&\sum\limits_{m=-\infty}^{\infty}\delta
\left(\phi_n-\theta_{n}-\epsilon\gamma\sum\limits_{k=0}^{n-1}|p_k|^2+2m\pi\right)
\\
&=&
\frac{1}{2\pi}\sum\limits_{m=-\infty}^{\infty}\exp\left[jm\left(\phi_n-\theta_n-\epsilon\gamma\sum\limits_{k=0}^{n-1}|p_{k}|^2\right)\right].
\end{IEEEeqnarray*}
where we have used phase periodicity and expanded the train of impulse
functions as a complex Fourier series. 

Summarizing, the conditional PDF now reads
\begin{multline}\label{eq:sumprod-p}
 f_{Q_n|Q_0}(q_n|q_0)=\frac{1}{2\pi}\sum\limits_{m=-\infty}^{\infty}e^{jm\phi_{n}}
\int\limits_{\theta_{n}=0}^{2\pi}e^{-jm\theta_{n}} \ \\ 
\frac{1}{\pi\sigma^2\epsilon}\int\limits_{\mathbb{C}} \ldots \int\limits_{ \mathbb{C}}
\exp\left\{-\epsilon\sum\limits_{k=0}^{n-1}\left[
\frac{1}{\sigma^{2}}\left|\frac{p_{k+1}-p_{k}}{\epsilon}\right|^{2}+j\gamma m|p_k|^2
 \right]\right\}\\
 \times\frac{dp_{n-1}}{\pi\epsilon\sigma^{2}}\frac{dp_{n-2}}{\pi\epsilon\sigma^{2}}\cdots\frac{dp_{1}}{\pi\epsilon\sigma^{2}},
\end{multline}
where we assumed $p_0=q_0$ and $p_n=r_n\exp j\theta_{n}$, and used the
fact that the Jacobian of the transformation \eqref{eq:changvar} is
unity, since it represents a lower triangular matrix with unit
magnitude on its diagonal elements.

The integrals in \eqref{eq:sumprod-p} are complex Gaussian and can be
solved directly. We proceed to calculate
\begin{IEEEeqnarray}{rCl}\label{eq:Tmn}
&&T_m(r_n)=\int\limits_{0}^{2\pi}d\theta_n e^{-jm\theta_n}\int\limits_{\mathbb{C}} \ldots \int\limits_{ \mathbb{C}}
\frac{dp_{n-1}}{\pi\epsilon\sigma^{2}}\frac{dp_{n-2}}{\pi\epsilon\sigma^{2}}\cdots\frac{dp_{1}}{\pi\epsilon\sigma^{2}}\times
\nonumber \\
&&\frac{1}{\pi\sigma^2\epsilon}\exp\left\{-\epsilon\sum\limits_{k=0}^{n-1}\left[
\frac{1}{\sigma^{2}}\left|\frac{p_{k+1}-p_{k}}{\epsilon}\right|^{2}+j\gamma m|p_k|^2
 \right]\right\}.
\end{IEEEeqnarray}

This is done by first integrating over phases, starting from the very
last term $\theta_{n}$ where only one variable is involved. It can be
shown that \eqref{eq:Tmn} simplifies to
\begin{IEEEeqnarray}{rCl}
&&T_m(r_n)=
\left(\frac{2}{\epsilon\sigma^2}\right)^n
e^{-jm(\theta_0+\gamma\epsilon r_0^2)}
\exp\left(-\frac{r_n^2+r_0^2}{\epsilon\sigma^2}\right)
\nonumber\\
&&
\int\limits_{r_{n-1}}^{\infty}\cdots\int\limits_{r_1}^{\infty}
dr_{n-1}\cdots dr_1
\nonumber\\
&&\prod\limits_{k=1}^{n-1}r_k\exp\left\{-\frac{2+jm\gamma \epsilon^2\sigma^2}{\epsilon\sigma^2}r_k^2\right\}
I_m\left(\frac{2r_kr_{k+1}}{\epsilon\sigma^2}\right)\\
&\times&
I_m\left(\frac{2r_1r_0}{\epsilon\sigma^2}\right),
\label{eq:Tmn2}
\end{IEEEeqnarray}
where $I_m$ denotes the $m^{th}$ order modified Bessel function of the
first kind, and where we have used \eqref{eq:Im} from
Appendix~\ref{sec:identities}. 

Integrals \eqref{eq:Tmn2} involve products of Bessel
functions and can be computed iteratively with the help of identity
\eqref{eq:Im1-Im2} in Appendix~\ref{sec:identities}. We have
\begin{IEEEeqnarray*}{rCl}
T_m(r_2)&=&\frac{2}{\epsilon\sigma^2(2+jm\gamma\epsilon^2\sigma^2)}e^{-jm(\theta_0+\gamma r_0^2\epsilon)}
\nonumber\\
&\times&
\exp\left\{-\frac{1}{\epsilon\sigma^2}\left(\frac{1+jm\gamma\epsilon^2\sigma^2}{2+jm\gamma\epsilon^2\sigma^2}\right)(r_0^2+r_2^2)\right\}
\nonumber\\
&\times&
I_m\left(\frac{2r_0r_2}{\epsilon\sigma^2 (2+jm\gamma\epsilon^2\sigma^2)}\right).
\end{IEEEeqnarray*}

As the calculations for $T_m(r_2)$ and identity \eqref{eq:Im1-Im2} suggest,
$T_m(r_n)$ keeps its structure for $n\geq 3$, and can be parametrized
in the form
\begin{IEEEeqnarray}{rCl}
\label{eq:Trn-form}
T_m(r_n)&=&2b_n(m) e^{-jm(\theta_0+\gamma r_0^2n\epsilon)}\exp\left\{-a_n(m)(r_0^2+r_n^2)\right\}
\nonumber\\
&\times&I_m\left(2b_n(m)r_0r_n\right),
\IEEEeqnarraynumspace
\end{IEEEeqnarray}
where $a_n(m)$ and $b_n(m)$ are parameters to be determined. It can be shown
that $T_m(r_n)$ satisfies
\begin{equation}
T_m(r_{n+1})=\int\limits_{r_n=0}^{\infty} K(r_n,r_{n+1})T_m(r_n)dr_n,
\label{eq:Tn-Tn-1}
\end{equation}
where kernel $K(r_{n+1},r_n)$ is 
\begin{IEEEeqnarray*}{rCl}
&&K(r_{n+1},r_n)=
\frac{2}{\epsilon\sigma^2}
r_n
\\
&&\exp\left\{-\frac{1}{\epsilon\sigma^2}\left[r_{n+1}^2+(1+jm\gamma\epsilon^2\sigma^2)r_n^2
  \right]\right\}
 I_m(\frac{2r_n r_{n+1}}{\epsilon\sigma^2}).
\end{IEEEeqnarray*}

Substituting \eqref{eq:Trn-form} into \eqref{eq:Tn-Tn-1} and comparing
exponents on both sides, recursive equations are obtained for $a_n(m)$
and $b_n(m)$, namely,
\begin{IEEEeqnarray*}{rCl}
b_{n+1}&=&\frac{b_n}{\epsilon\sigma^2[a_n+\frac{1+jm\gamma\epsilon^2\sigma^2}{\epsilon\sigma^2}]},
\\
a_{n+1}&=&\frac{1}{\epsilon\sigma^2}-\frac{1}{(\epsilon\sigma^2)^2[a_n+\frac{1+jm\gamma\epsilon^2\sigma^2}{\epsilon\sigma^2}]}
\\
&=&a_n-\frac{jm\gamma\epsilon^2\sigma^2}{\epsilon\sigma^2}
-\frac{b_n^2}{[a_n+\frac{1+jm\gamma\epsilon^2\sigma^2}{\epsilon\sigma^2}]},
\end{IEEEeqnarray*}
with
\begin{IEEEeqnarray*}{rCL}
a_2=\frac{1+jm\gamma\epsilon^2\sigma^2}{\epsilon\sigma^2(1+jm\gamma\epsilon^2\sigma^2)},
\qquad
b_2=\frac{1}{\epsilon\sigma^2(2+jm\gamma\epsilon^2\sigma^2)}.
\end{IEEEeqnarray*}
Solving these equations, we obtain expressions for $a(m)$ and $b(m)$
in the limit as $n\rightarrow\infty$ ($n\epsilon=\const{L}$):
\begin{IEEEeqnarray}{rCl}\label{eq:anbn}
a(m)&=&\lim_{\substack{n\epsilon=\const{L}\\
n\rightarrow \infty }}a_n(m)=\frac{\sqrt{jm\gamma }}{\sigma}\coth\sqrt{j m\gamma \sigma^2 }\const{L},\nonumber \\
b(m)&=&\lim_{\substack{n\epsilon=\const{L}\\
n\rightarrow \infty }}b_n(m)=\frac{\sqrt{jm\gamma}}{\sigma}\frac{1}{\sinh\sqrt{jm\gamma\sigma^2}\const{L}}.
\end{IEEEeqnarray}

Finally, using \eqref{eq:anbn} we obtain the conditional PDF of $Q$ in
the zero-dispersion model \eqref{eq:zerodis}
\begin{IEEEeqnarray}{rCl} \label{eq:pdf}
f(r,\phi|r_{0},\phi_{0})&=&\frac{f_{R|R_{0}}(r|r_{0})}{2\pi}\nonumber \\
&&\:+\frac{1}{\pi}\sum_{m=1}^{\infty}\Re\left(C_{m}(r)e^{jm(\phi-\phi_{0}-\gamma r_0^2\const{L})}\right),\IEEEeqnarraynumspace
\end{IEEEeqnarray}
where $f_{R|R_{0}}(r|r_{0})$ is the probability density of the
amplitude of the received signal
\begin{equation} \label{eq:ampdf}
  f_{R|R_{0}}(r|r_{0})=\frac{2r}{\sigma^{2}\const{L}}\exp(-\frac{r^{2}+r_{0}^{2}}{\sigma^{2}\const{L}}) I_{0}(\frac{2rr_{0}}{\sigma^2\const{L}}),
\end{equation}
and where the Fourier coefficient $C_{m}(r)$ is given by
\begin{IEEEeqnarray*}{rCl}
C_m(r)= r b(m)\exp\left[-a(m)\left(r^2+r_0^2\right)\right]I_{m}(2b(m)r_0 r).\IEEEeqnarraynumspace
\end{IEEEeqnarray*}

\begin{remark}
  \label{rem:symmetry}
Note that $f(r,\phi|r_{0},\phi_{0})$ is symmetric with respect to the
phase $\phi$. We rely on this property in Sec.~\ref{sec:num-eval} to
simplify the optimization problem in the capacity question.
\end{remark}
\begin{remark}
\label{rem:imdd}
One can verify that the conditional PDF of the amplitude of the
received field given the amplitude of the transmitted signal
\eqref{eq:ampdf} is in fact the conditional PDF for the
intensity-modulated direct-detection (IM/DD) channel
\begin{IEEEeqnarray}{rCl}\label{eq:imdd}
R_k=\left|Q_{0}^{k}+Z_k\right|\quad k=0,1,\ldots,
\end{IEEEeqnarray}
where  $Z_{k}\sim \mathcal{N}_{\mathbb{C}}(0,\sigma^{2}\const{L})$. It
is easy to see that the conditional PDF depends only on $R_0=|Q_0|$,
as in (\ref{eq:ampdf}).
\end{remark}

The conditional PDF (\ref{eq:pdf})
defines a communication channel
having the complex plane as
the input alphabet,
for which the
information capacity $\const{C}$ is defined as
\begin{IEEEeqnarray}{rCl}\label{persample-c}
\const{C}(\const{P},\sigma^{2}\const{L},\gamma)&=&\max\limits_{f(r_{0},\phi_{0})\in\const{F}}\quad
I(R,\Phi;R_{0},\Phi_{0}) \nonumber \\
&&\textnormal{subject to}\quad ER_{0}^{2}\leq \const{P},
\end{IEEEeqnarray}
where
\begin{IEEEeqnarray*}{rCl}
I(R,\Phi;R_{0},\Phi_{0})&{=}& \iiiint f(r_{0},\phi_{0})f(r,\phi|r_{0},\phi_{0})\nonumber \\
&&
\cdot\:\log\frac{f(r,\phi|r_{0},\phi_{0})}{f(r,\phi)}dr_0
d\phi_0 dr d\phi,\IEEEeqnarraynumspace
\end{IEEEeqnarray*}
and where $\const{F}$ is the space of
probability densities, and $f(r,\phi)=\iint
f(r,\phi|r_{0},\phi_{0})f(r_{0},\phi_{0})dr_{0}d\phi_{0}$.

The per-sample capacity (\ref{persample-c}) can be related to the
capacity of the waveform dispersion-free optical channel. It can be
argued that a uniform power allocation is optimal for principal samples.
The capacity of the entire ensemble of the principal per-sample
channels is
\begin{equation*}
2\const{W}_{0}\const{T}\const{C}\left(\frac{\const{P}_0}{2\const{W}_{0}\const{T}},\frac{2\sigma_{0}^{2}
\const{L}\const{W}_{\const{L}}}{2\const{W}_{0}\const{T}},\gamma\right)
\quad \textnormal{bits per channel use},
\end{equation*}
which, as discussed before, is a lower bound on the capacity of the
zero-dispersion waveform channel (\ref{eq:zerodisw}). Therefore in
this paper we only estimate the per-sample capacity of
(\ref{eq:zerodis}).

The following theorem is a simple way to establish the results of
\cite{turitsyn2003ico}.
\begin{theorem}
  \label{thm:asymp-c}
Let $\rho=\const{P}/\sigma^2\const{L}$ be the
signal-to-noise ratio. Then $\const{C}\geq \frac{1}{2}\log
\rho-\frac{1}{2}$
and in particular $\lim \limits_{\rho \rightarrow \infty}\const{C}(\rho)=\infty$.\\
\end{theorem}
\begin{IEEEproof}
Using the chain rule for mutual information
\begin{IEEEeqnarray*}{rCl}
 I(R,\Phi; R_{0},\Phi_{0})&=&I(R;R_{0},\Phi_{0})+I(\Phi;R_{0},\Phi_{0}|R) \nonumber \\
&\geq& I(R;R_{0},\Phi_{0})=I(R;R_{0}).
\end{IEEEeqnarray*}

As mentioned in \remref{rem:imdd}, $I(R;R_{0})$ is the mutual
information function for the intensity-modulated direct-detection
(IM/DD) channel for which the lower bound  $\frac{1}{2}\log
\rho-\frac{1}{2}$ is already known \cite{mecozzi1994llh}. It therefore
follows that $\const{C} \geq \frac{1}{2}\log \rho-\frac{1}{2}$.
\end{IEEEproof}

Put in other words, \thmref{thm:asymp-c} simply says that the amount of
information which can be sent over the complex channel
\eqref{eq:zerodis} is no less than what can be transmitted and
received by the amplitude alone. From \eqref{eq:ampdf}, the
communication channel from $R_0$ to $R$ does not depend on the
nonlinearity parameter $\gamma$, is independent of input phase and
supports an unbounded information rate when increasing power
indefinitely. Since the capacity of \eqref{eq:zerodis} was discussed
to be a lower bound to the capacity of \eqref{eq:zerodisw}, we
conclude that the capacity of the zero dispersion optical fiber
\eqref{eq:zerodisw} also goes to infinity with \SNR. 

\subsection{Sum-product probability flow in zero-dispersion fibers}
\label{subsec:sumproduct}

The recursive computation of the PDF in the previous section was
algebraic and still not a suitable way to visualize signal statistics.
In this section, we show that the structure of the probability flow in
zero-dispersion fibers is given by the sum-product algorithm or a path
integral. Such path integration underlies the Martin-Siggia-Rose
formalism in quantum field theory (QFT), which was directly used in
\cite{turitsyn2003ico}. This connects the results of the previous
section to \cite{turitsyn2003ico}. 

Computation of the conditional PDF as explained
in the previous section was a marginalization process. In our
example, the Markov property \eqref{markov} made it possible to
perform marginalization, since the conditional PDF factors as a
product of certain normalized functions. This observation 
allows us to apply the
\emph{sum-product algorithm} known already in coding theory. As a
matter of fact, the reader can notice that \eqref{eq:pathint1-1} is a
sum-product computation, with a slight difference that instead of
multiple summation, we performed multiple integrations, which is
indeed a continuous limit of the sum-product algorithm in the signal
dimension. It might be, however, more insightful to restore the
analysis and represent the technique in the discrete domain.

\begin{figure*}[tbp]
\centerline{\includegraphics[width=\textwidth]{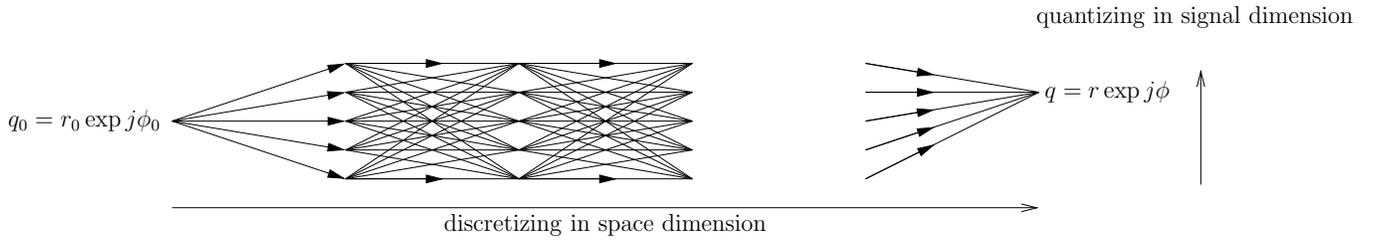}}
\caption{Graphical computation of the probability density function
$f(r,\phi | r_{0}, \phi_{0})$. The probability of receiving
$r_n\exp\phi_m$ given transmission of $r_{k}\exp\phi_{l}$ is the sum
over the probability of each possible path between start and end
points shown schematically in the figure.  The probability of each
path is the product of of the transition probabilities of all edges
forming that path, computable from (\ref{incdens}). }
\label{fig:path-integ}
\end{figure*}

While discretizing the fiber in the distance dimension, we will
also at
the end of at each fiber segment quantize the signal $q_k$ into a
large number of small bins in the complex plane
\begin{IEEEeqnarray}{rCl}\label{eq:quantization}
q_k \in \set{S}&=&\{q_n^m=r_n\exp(j\phi_m)|r_n=n(\delta r),\quad
n=1,\ldots,N,\quad\nonumber \\
&&\phi_m=m(\delta\phi),\quad m=0,1,\ldots,M-1 \},\IEEEeqnarraynumspace
\end{IEEEeqnarray}
for small $\delta r$ and $\delta\phi$. This turns each incremental channel
into a discrete memoryless channel (DMC) described by a transition
matrix; moreover, the overall channel matrix is the product of all of these
transition matrices. The probability of receiving $q_k^l=r_k\exp
j\phi_l$ at $z=\const{L}$ given the $q_i^j=r_i\exp j\phi_j$ is
transmitted at $z=0$, is the sum over the probability of all possible
transitions (paths) from $q_i^j$ to $q_k^l$. This is graphically
illustrated in Fig.~\ref{fig:path-integ} as a trellis. Nodes of the
trellis are quantized points in the complex plane and edges have
weights corresponding to transition probabilities \eqref{incdens}.

It follows that the essence of the probabilistic model in optical
fibers is a \emph{propagator} $\matd{A}(\delta
r,\delta\phi,\epsilon)$, independent of spatial index $k$, which for
the case of \eqref{eq:zerodis} can be described by a single functional
matrix
\begin{IEEEeqnarray}{rCl}\label{eq:inc-chmat}
[\matd{A}(\delta r,\delta\phi,\epsilon)]_{i,j}= \frac{|q_n^m|\delta r\delta \phi}{\pi \sigma^{2}\epsilon}
\exp\left(-\frac{\left|q_n^m-q_k^l-j\epsilon\gamma
q_k^l|q_k^l|^{2}\right|^2}{\epsilon
 \sigma^{2}}\right),\nonumber
\end{IEEEeqnarray}
where $i=(k-1)M+l$ and $j=(n-1)M+m$. The probabilities of
$\{q_n^m\}_{n,m=0}^{N,M}$ are then recursively updated as
\begin{IEEEeqnarray}{rCl}
\vecd{P}(k+1)=\matd{A}(\delta r,\delta\phi,\epsilon)\vecd{P}(k),\nonumber
\end{IEEEeqnarray}
where $\vecd{P}$ is the vector of the probabilities of
$\{q_n^m\}_{n,m=0}^{N,M}$. By diagonalizing the propagator
$\matd{A}=\matd{U}\matd{\Lambda}\matd{U}^{-1}$, the overall
conditional distribution is
$\vecd{P}(n)=\matd{U}\matd{\Lambda}^{n}\matd{U}^{-1}$, which is a
function of eigenvalues and eigenvectors of the propagator.

In the language of the statistical mechanics, the limit of the
expression \eqref{eq:pathint1-2} when $n\rightarrow\infty$ is a
\emph{path integral} and is represented as
\begin{IEEEeqnarray}{rCl}\label{eq:pathint}
&&f_{Q|Q_{0}}(q|q_{0})=\nonumber \\
&&\int_{  r_{0}e^{j\phi_{0}} }^{re^{j\phi}}\exp\left(-\int_{0}^{z}\frac{\left|\frac{\partial
q}{\partial
z^{\prime}}-j\gamma|q(z^{\prime})|^{2}q(z^{\prime})\right|^{2}}{\sigma^{2}}dz^{\prime}\right)\mathfrak{D}q,\IEEEeqnarraynumspace
\end{IEEEeqnarray}
where the expression is understood as in \cite{feynman1965}.
Equation (\ref{eq:pathint}) is just a symbolic representation of
\eqref{eq:pathint1-2} and follows from the definition of the path
integral \cite{feynman1965}, \cite{zinnjustin2002qft}.

The
computation of path integrals whose exponent can be made quadratic in
$q$ and $\partial q/\partial z$ is standard in quantum mechanics and,
in the case of (\ref{eq:pathint}), this computation
has been done in
\cite{turitsyn2003ico} to find the conditional PDF.
Indeed
path integral \eqref{eq:pathint} is an immediate consequence of the
Martin-Siggia-Rose formalism, a more generic framework in quantum
mechanics dealing with stochastic dynamical systems
\cite{turitsyn2003ico}.
To compute (\ref{eq:pathint}), roughly
speaking, one needs to sum over the input-output paths giving
the largest
contributions, \emph{i.e.}, the path
corresponding to minimizers of the integral exponent in
(\ref{eq:pathint}).  This classical
path is given by the Euler-Lagrange equations,
with all other paths considered as a perturbation around this classical
path, serving as a normalization constant in the PDF. 

It is interesting to see the effect of the change in variables
\eqref{eq:changvar} in the sum-product algorithm. One can verify that
the equation \eqref{eq:zerodis} when noise is zero has the solution
\begin{IEEEeqnarray}{rCl}\label{eq:zdsol}
q(z)=q_0\exp(j\gamma|q_0|^2z).
\end{IEEEeqnarray}
In other words, while the signal is propagated down the noiseless fiber,
it remains on a circle in the complex plane with radius $|q_0|$ and
rotates counterclockwise with phase velocity $\gamma |q_0|^2$ rad/m.
We call this solution the \emph{deterministic path}, which is a helix
twisted around the fiber. With the change of variable
\eqref{eq:changvar}, the accumulated phase from the beginning of the
fiber to an arbitrary distance $z$ is subtracted from the (nonlinear)
phase of the signal, to compensate for the signal rotation. The
deterministic path is then a straight line, rather than a twisted
helix. In this view, \eqref{eq:zdsol} gives rise to
nonlinear phase
compensation at the receiver, often used in optical communications.

\begin{figure*}[tbp]
\centerline{\includegraphics[width=\textwidth]{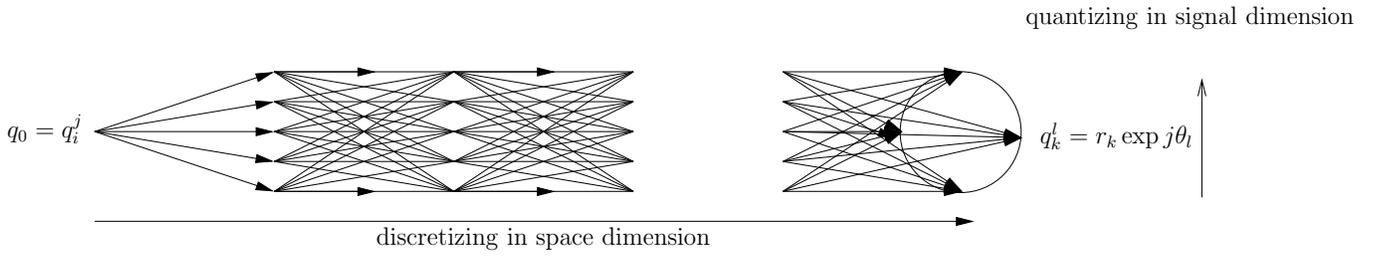}}
\caption{Transformation of the q-trellis (the trellis in Fig.~\ref{fig:path-integ}) under the transformation (\ref{eq:changvar}).
The result is called the p-trellis.} \label{fig:p-trellis}
\end{figure*}

When using \eqref{eq:changvar}, the trellis in
Fig.~\ref{fig:path-integ} is transformed to a new trellis. Since the
change of variable \eqref{eq:changvar} has memory of the form $\exp
\left(-j\epsilon\gamma
\sum\limits_{m=0}^{k-1}\left|q_m\right|^2\right)$, it causes a
transformation from paths to points
\begin{IEEEeqnarray*}{rCl}
\mathbb{P}:(q_0,q_1^{m_1},q_2^{m_2},\cdots,q_k^{m_k})\mapsto p_k^l.
\end{IEEEeqnarray*}
In general $q_n^m$ at stage $k$ of the old trellis is mapped to a
number of points at stage $k+1$ in the new trellis, depending on the
total number of paths from $q_0$ to $q_n^m$. These points lie on a
circle and correspond to rotations of constellations in the old
trellis. The structure of the $p$-trellis, in the limits $\delta
r,\delta\phi\rightarrow 0$, is same as Fig.~\ref{fig:path-integ}
except that instead of a single terminal point at the end of the
diagram, there is a circle of radius $|q_k^l|$. The transformed
trellis is shown in Fig.~\ref{fig:p-trellis}.

The expression \eqref{eq:pathint1-1} is then the sum over the
probability of all transition paths starting with $q_0=q_i^j$ and
ending in any of the resulting terminal points. The overall sum can be
decomposed into a number of subgroups, with each subgroup a sum-product from
$r_i\exp j\phi_j$ to one of the output terminal points $r_k\exp
j\theta_l$, for some $l$ as in \eqref{eq:quantization}. The total sum
is then obviously sum over all these sub-sum-products corresponding to
different phase levels
\begin{multline*}
P[r_k\exp j\phi_l|r_i\exp j\phi_j]\\
=\sum\limits_{ \textnormal{all possible }\theta_l}\mathbb{\textnormal{SP}}(q_0=q_i^j\rightarrow q_k^l=r_k\exp j\theta_l),
\end{multline*}
where $\mathbb{\textnormal{SP}}(q_0=q_i^j \rightarrow q_k^l=r_k\exp
j\theta_l)$ is a sum-product from $q_i^j$ to $q_k^l$. In terms of
Fig.~\ref{fig:p-trellis}, this is a sum-product from $q_0$ to one of
the points on the terminal circle. Note that, however, each
sum-product from  $r_i\exp j\phi_j$ to $r_k\exp j\theta_l$ is a
\emph{constrained} sum-product, \emph{i.e.}, instead of all possible
paths between these two points, the summation in \eqref{eq:Tmn} is
performed over only feasible paths consistent with
\eqref{eq:phase-const}.  This is accomplished by
multiplying the edge-weights
by an \emph{indicator function}
representing \eqref{eq:phase-const}, and then summing over all possible
paths between $r_i\exp j\phi_j$ to $r_k\exp j\theta_l$. The indicator
function is later expanded in terms of Fourier series to make
analytical computations possible.  

\section{Numerical Evaluation of the Capacity}\label{sec:num-eval}

In this section we numerically evaluate the per-sample capacity of the
communication channel (\ref{eq:zerodis}) to observe the general trend
of the capacity as a function of the input average power. In
particular, we are interested in observing the effect of the
signal-dependent noise in the absence of dispersion. The spectral
efficiency of the dispersive optical channel as a function of the
input power is known to have a peak \cite{mitra2000nli}. The peak is
often attributed to the fact that increasing the signal power will
increase the noise power as well (which is signal-dependent). The same
type of behavior was observed in \cite{tang2001scc} for the
nondispersive channel as well. In this section, we numerically observe
that with no dispersion, although the channel is still nonlinear, the
signal-dependent noise is not strong enough to suppress the capacity
to zero.

We consider a 5000 km optical fiber operating at zero average
dispersion and using distributed Raman amplification
\cite{islam2003rat}. Among several sources of noise, ASE noise is
assumed to be the dominant stochastic impairment, which can be modeled
as additive white Gaussian noise. All other nominal simulation
parameters are given in Table.~\ref{tbl:fiberparam}
\cite{essiambre2010clo, agrawal2001nfo}.

We sample the conditional PDF (\ref{eq:pdf}) in a high resolution grid
in the complex plane with $N$ rings and $M$ symbols on each ring.  The
values of $N$ and $M$, or the size of bins, depend on the noise standard
deviation and are chosen so that in each noise standard deviation
there are sufficiently many bins so that the conditional PDF of the
channel is normalized over the entire partition. This is very similar
to the multiple ring modulation format, an idea well-known in the
context of AWGN channels and recently employed in optical
communication in \cite{essiambre2010clo}.

The accumulated noise level in the optical fiber, albeit signal
dependent, is still much smaller than the noise level in typical AWGN
channels. Thus compared to the AWGN channels, more resolution is
needed to numerically approach the capacity of optical channels via
this technique. In addition, it is known that in a dispersive channel,
dispersion and nonlinearity to some extent cancel out each some of the
detrimental effects of the other, leading to a balanced propagation.
In the absence of dispersion such balance does not exist anymore, and
the channel exhibits stronger nonlinear properties. This makes it more
difficult to numerically evaluate the capacity of the dispersionless
channel. 

Estimates of the capacity are found and compared using two methods:
the Blahut-Arimoto algorithm with power constraint
\cite{blahut1972ccc} and a logarithmic barrier interior-point method
\cite{boyd2004cop}. From the phase symmetry of the channel matrix, a
capacity-achieving input distribution will have uniform phase
distribution. In both algorithms, we enforce this constraint to make
the problem effectively a one dimensional optimization program, which
considerably stabilizes and speeds up the underlying numerical
optimization.

Note that the Blahut-Arimoto algorithm in its original form was
developed for linear inequality constraints, and cannot be directly
used to approximate the capacity in a region where $\const{C}$ is
decreasing with $\const{P}$. One might modify the original
Blahut-Arimoto algorithm to take into account the equality constraints
as well. However since in our numerical simulations capacity is never
found to be constant when increasing power, the inequality power
constraint is always active at the optimal solution and no such
modification is required. Because for small \SNR s capacity follows
Shannon's limit for linear Gaussian channels, the Lagrange
multiplier is started at a large value and is iterated down to zero at
the optimum, with increasing resolution.

\begin{figure}[tbp]
\centerline{\includegraphics[width=0.95\columnwidth]{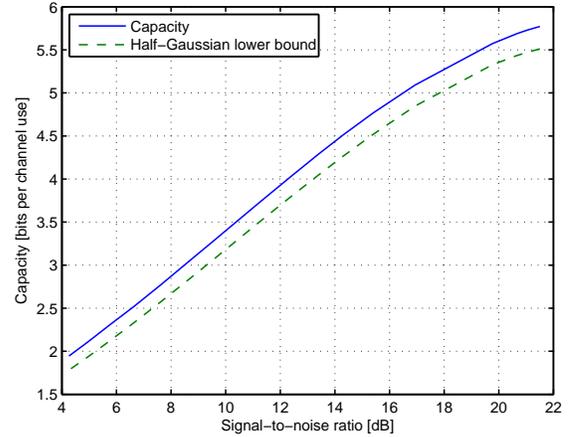}}
\caption{Capacity as a function of SNR with peak power
$\const{P}_{0}=10$mW and noise spectral density $10^{-1}\mu$W/GHz. The
end of the graph occurs at about 20\% of the peak power.}
\label{fig:CvsP2}
\end{figure}

\begin{figure}[tbp]
  \centering
  \subfigure[]{\label{fig:phase:subl}\includegraphics[width=0.95\columnwidth]{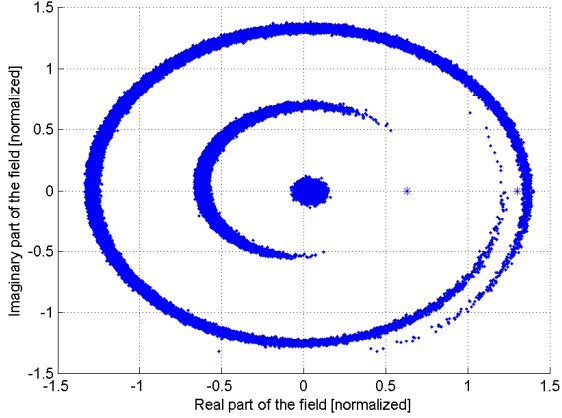}}
  \subfigure[]{\label{fig:phase:sub2}\includegraphics[width=0.95\columnwidth]{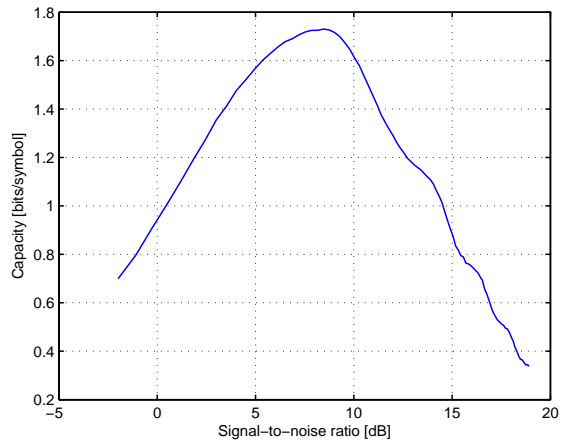}}
  \caption{(a) Simulation of the optical phase channel channel
operating at zero-dispersion for 3 transmitted points. It can be seen
that for signals with very low or high power, the phase of the
received signal contains almost no information.  (b) Capacity of the
phase modulation dispersion-free optical channel as a function of
SNR.}
  \label{fig:phase}
\end{figure}

\begin{figure}[tbp]
\centerline{\includegraphics[width=0.95\columnwidth]{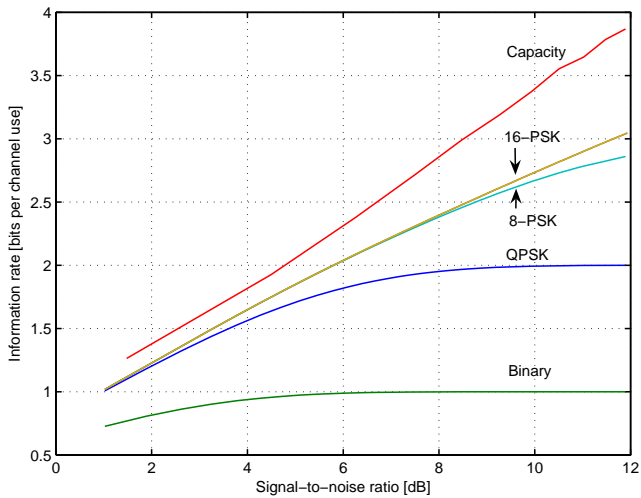}}
\caption{Achievable information rates in zero-dispersion optical fiber with
  M-PSK modulation formats} \label{fig:mpsk}
\end{figure}

The results of the numerical calculation of the capacity are shown in
Figs.~\ref{fig:CvsP2}-\ref{fig:mpsk}. Fig.~\ref{fig:CvsP2}
shows the per-sample capacity $\const{C}$ of channel
\eqref{eq:zerodis} as a function of input average signal power
$\const{P}$ for noise spectral density $10^{-1}~\mu$ W/GHz. We can see
that the channel capacity increases indefinitely with the average input 
power. Moreover, as evident in Fig.~\ref{fig:CvsP2}, in the
high power regime the capacity growth is linear on a logarithmic
scale. To simulate at increased power levels, a prohibitive number
of symbols is required.

Motivated by previous work \cite{tang2001scc,turitsyn2003ico}, we
searched for peaks in the capacity curve for different set of fiber
parameters. Simulations were performed for a wide range of parameters to
operate at low and high \SNR s. No peak was found within our
computational ability of finding capacity at high \SNR s.

Although there is no peak in the capacity curve, the phase channel
$\Phi_{0} \mapsto (R,\Phi)$ was found to always having a global
maximum in its $\const{C}-\const{P}$ plot. Intuitively this behavior
is a consequence of the observation that the phase of the received field is
uniform and independent of the transmitted signal, if very low or high
energy signals are sent through the channel. For a fixed noise
variance, when signal power level $\const{P}$ is
small, phase supports little information because the linear phase
noise dominates and is uniformly distributed in $[0,2\pi]$. On the
other hand
when $\const{P}$ is large, the nonlinear phase noise takes over the
whole interval [0,$2\pi$] and the phase sub-channel is
unable to carry  data. This
effect is illustrated in Fig.~\ref{fig:phase}.

In Fig.~\ref{fig:mpsk} we have plotted the capacity and the mutual
information that uncoded phase-shift keying modulation formats can
achieve in the zero-dispersion optical fiber. Similar to the AWGN
channel, binary signaling is suboptimal at low SNRs (\emph{e.g.},
$\textnormal{SNR} < 0$). More symbols are required to get close to the
capacity with increasing SNR. 

\section{An Algebraic Model}\label{sec:simmod}

Although we used the PDF (\ref{eq:pdf}) to numerically find the
capacity, it provides limited direct information-theoretic insights
into the behavior of the channel.
We therefore proceed with a more intuitive and
simplified expression for the channel model. We reduce the
\emph{differential model} \eqref{eq:zerodis} to an \emph{algebraic
model}, which is more tractable for an information-theoretic
analysis. 
\begin{figure}[tbp]
\centerline{\begin{tabular}{c c}
\includegraphics*[width=0.49\columnwidth]{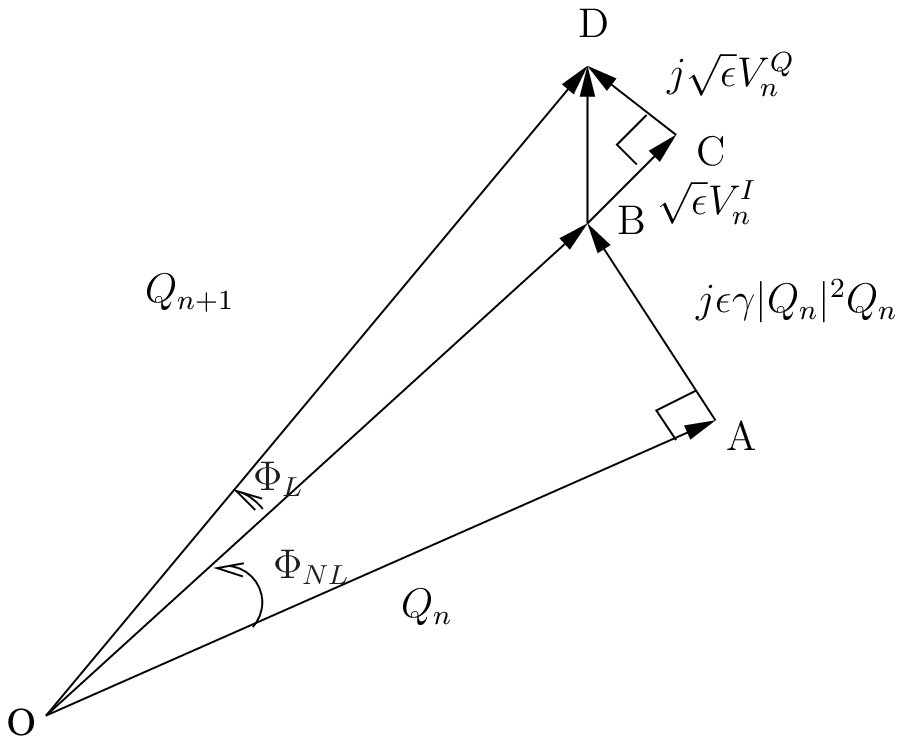} &
\raisebox{1cm}{\includegraphics*[width=0.49\columnwidth]{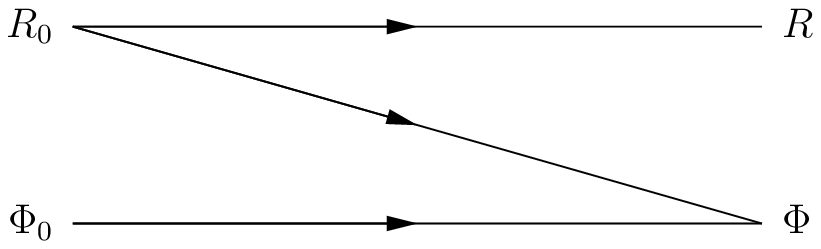}} \\
(a) & (b)
\end{tabular}}
  \caption{(a) Vector diagram showing the signal evolution in the $n^{\textrm{th}}$ incremental piece of the fiber
  (b) Information streaming in the dispersion-free optical channel. Note that no information is transferred from $\Phi_{0}$ to $R$.}
\label{fig:vec-diagrams}
 \end{figure}

The vector diagram in Fig.~\ref{fig:vec-diagrams}(a) pictorially shows
the evolution of the signal in the $n^{\textrm{th}}$ incremental piece
of the fiber. The nonlinear term $j|Q|^2Q$ is orthogonal to the signal
and, to the first order in $\epsilon$, does not change the amplitude.
In other words, since in Fig.~\ref{fig:vec-diagrams}(a)
\begin{IEEEeqnarray*}{rCl}
|\vec{OB}|=|\vec{OA}|+O(\epsilon^{2}),
\end{IEEEeqnarray*}
the orthogonality of the nonlinearity to the signal implies that the
nonlinear coefficient $\gamma$ is responsible only for an angular
rotation; it does not contribute to random fluctuations across the
radius.
Therefore for the description of
the amplitude channel we can set $\gamma=0$ without incurring any
local or global error at the end of the fiber
\begin{IEEEeqnarray}{rCl} \label{eq:recamp}
 |Q_{n+1}|= |\vec{OD}|&=&|\vec{OB}+\sqrt{\epsilon} V_{n}^{I}+j\sqrt{\epsilon} V_{n}^{Q}|\nonumber \\
&=&|Q_{n}+\sqrt \epsilon V_{n}|+O(\epsilon^2),
\end{IEEEeqnarray}
where $V^I$ and $V^Q$ are in-phase and quadrature components of the
noise, and we have used the fact that white noise preserves its
properties under rotation. Similarly for the phase channel
\begin{IEEEeqnarray}{rCl} \label{eq:recphase}
\Phi_{n+1}&=&\Phi_{n}+\tan^{-1}j\gamma \epsilon |Q_{n}|^2+\tan^{-1}
\frac{|\sqrt\epsilon V_{n}^{Q}|}{|R_{n}+\sqrt \epsilon
V_{n}^{I}|}\nonumber \\
&=&\Phi_{n}+\Phi_{L}+j\gamma\epsilon |Q_{n}|^{2}+O(\epsilon^2).
\end{IEEEeqnarray}
Hence from (\ref{eq:recamp})-(\ref{eq:recphase}) it follows that
\begin{equation} \label{rse1}
 R(z)=|Q_{0}+W(z)|=|R_{0}+W(z)|,
\end{equation}
\begin{equation} \label{phise1}
\Phi(z)=\Phi_{L}(z)+\gamma \int_{0}^{z} |Q_{0}+W(z)|^2,
\end{equation}
where $W(z){=}\int_{0}^{z}V(\lambda)d\lambda$ is the Wiener process
with $EW(z)W^*(z^{\prime}){=}\sigma^{2}\min(z,z^{\prime})$. The linear
part of the phase $\Phi_{L}(z)$ from Fig.~\ref{fig:vec-diagrams}(a) is
only a function of the amplitude $|\vec{OB}|$ and the noise $V_{n}$.
As a result, setting $\gamma=0$ in (\ref{eq:zerodis}) it is given by
\begin{equation} \label{phiL}
 \Phi_{L}(z)=\angle (Q_{0}+W(z)).
\end{equation}

A byproduct of (\ref{rse1})-(\ref{phiL}) is the compact solution of
the nonhomogeneous stochastic ODE (\ref{eq:zerodis})
\begin{IEEEeqnarray}{rCL}
\label{eq:stoch-sol-zd}
Q(z)&=&(Q_{0}+W(z))\exp(j \gamma
\int\limits_{0}^{z}|Q_{0}+W(z^{\prime})|^{2} dz^{\prime}).
\end{IEEEeqnarray}

The first equation in the system (\ref{rse1})-(\ref{phise1}) is
decoupled from the second one and hence, neither deterministic
amplitude nor its noisy perturbation depends on the Kerr nonlinearity
constant $\gamma$. In particular, from (\ref{rse1}) the probability
density function of $R$ follows, as in (\ref{eq:ampdf}). This is
schematically shown in Fig.~\ref{fig:vec-diagrams}(b).
Information is streamed from $R_{0}$ to $(R,\Phi)$, but $\Phi_{0}$
does not excite the output amplitude.

Amplitude and phase channels are defined by (\ref{rse1})-(\ref{phiL}).
Statistics can be directly computed from these equations and are
generally signal-dependant Gaussian noises, and Gaussian squared
(chi-squared) noise-noise beats. The original PDF can be rederived by
calculating these statistics. To simplify the model, we replace the
noise-noise beat terms with worst case Gaussian noises with the same
mean and covariance matrix. In addition, we ignore the linear phase
$\Phi_L$ compared to $\Phi_{\textnormal{NL}}$, which is stochastically
valid for $\const{P}\gg \sigma^2\const{L}$. Though it is possible to
calculate the statistics directly, it is generally easier to replace
the Wiener process with its Karhunen-Lo\'{e}ve (KL) expansion in order
to substitute the correlated sums with summations with uncorrelated
terms and ease the following calculations. We are in accord of
\cite{ho2003pdn} concerning the use of Karhunen-Lo\'{e}ve expansion
for such simplification. The KL expansion for the Wiener process reads
\cite{gardiner1985hsm}
\begin{equation*}
 W(z)=\sum\limits_{k=1}^{\infty} \sigma_{k}X_{k}\psi_{k}(z), \qquad
 0\leq z\leq \const{L},
\end{equation*}
where $\{X_{k}\}\sim\textrm{IID}$
$\mathcal{N}_{\mathbb{C}}(0,\sigma^{2}\const{L})$  is a sequence of
independent identically distributed zero-mean complex Gaussian random
variables, and eigenvalues $\sigma_{k}$ and eigenfunctions
$\psi_{k}(z)$ are defined as
\begin{IEEEeqnarray*}{rCl}
\psi_{k}(z)&=&\sqrt{2}\sin\left(\frac{(2k-1)\pi}{2\const{L}}z\right), \qquad 0\leq z\leq \const{L},\\
\sigma_{k}&=&\frac{2}{(2k-1)\pi}.
\end{IEEEeqnarray*}

At the end of the fiber $z=\const{L}$ we have, after some algebra,
\begin{IEEEeqnarray}{rCl} \label{chmod0}
R^2& =& |Q_{0}+Z_{1}|^{2}=R_{0}^{2}+2\Re{\{Q_{0}Z_{1}\}}+|Z_{1}|^{2}, \nonumber\\
\Phi & \approx & \Phi_{0}+\gamma\const{L}
\left(R_{0}^{2}+\frac{2}{\sqrt{3}}\Re{\{Q_{0}Z_{2}\}}+Z_{3}\right),
\end{IEEEeqnarray}
where $Z_{1}\sim \mathcal{N}_{\mathbb{C}}(0,\sigma^{2}\const{L})$,
$Z_{2}\sim \mathcal{N}_{\mathbb{C}}(0,\sigma^{2}\const{L})$,
$EZ_{1}Z_{2}^{*}=\frac{\sqrt{3}}{2}\sigma\sqrt{\const{L}}$ and
$Z_{3}=\sum_{k=0}^{\infty}\sigma_{k}^{2}|X_{k}|^{2}$ is a
non-Gaussian random variable correlated with $Z_{1}$ and $Z_{2}$.
In matrix notation
\begin{equation} \label{eq:sim-model}
 \vecr{Y}\simeq\vecr{X}+\matr{H}\vecr{Z}_{1}+\vecr{Z}_{2}, \quad
 \vecr{Y}_{2} \: \textrm{mod}\:
2\pi,
\end{equation}
in which
\begin{equation*}
 \vecr{Y}=\left(\begin{array}{ll}
R^{2}   \\
\frac{\Phi}{\gamma\const{L}}
 \end{array}\right), \
\vecr{X}=\left(\begin{array}{ll}
R_{0}^{2}   \\
\frac{\Phi_{0}}{\gamma\const{L}}+R_{0}^{2}
 \end{array}\right),
\mathbb{H}=\left(\begin{array}{ll}
 R_{0}& 0 \\
0 & R_{0}
 \end{array}\right),
\end{equation*}
 and $\vecr{Z}_{1}\sim\normalr{0}{\matd{P}_{1}}$, while $\vecr{Z}_{2}$
is a non-zero mean random variable related to central chi-square
random variables with
$\E(\vecr{Z}_{2}-\E\vecr{Z}_{2})(\vecr{Z}_{2}-\E\vecr{Z}_{2})^{T}=\matd{P}_{2}$.
It can be shown that
\begin{equation*}\label{eq:covmat}
 \matd{P}_{1}=\sigma^{2}\const{L}\left(\begin{array}{ll}
 2 & 1 \\
1 & \frac{2}{3}
 \end{array}\right), \quad\matd{P}_{2}=\sigma^{4}\const{L}^2\left(\begin{array}{ll}
 1 & \frac{1}{3} \\
\frac{1}{3} & \frac{1}{6}
 \end{array}\right),\quad \vecr{Z}_1\perp\vecr{Z}_2.
\end{equation*}
Gaussian random variables in model (\ref{eq:sim-model}) can be further
decoupled by performing the Cholesky factorization of the
$\matd{P}_{1}=LL^{T}$ and processing $\matd{L}^{-1}\vecr{Y}$ as the
output. Equations \eqref{chmod0} can also be approximated as
\begin{IEEEeqnarray*}{rCl}
R&=&|R_0+Z_1|,\nonumber \\
\Phi&\approx&\Phi_0+\gamma\const{L}|R_0+\frac{Z_2}{\sqrt{3}}|^2,\qquad \const{P}\gg \sigma^2\const{L}.
\end{IEEEeqnarray*}

From (\ref{eq:sim-model}), nonlinearity introduces two important
stochastic effects which are always absent in linear channels:
signal-noise beat $\matd{H}\vecr{Z}_1$ and non-Gaussian noise-noise
beat $\vecr{Z}_2$. In our problem, signal-noise beat is a simple
signal-dependent noise in the form of the product of a Gaussian random
variable with amplitude of the signal and stands for the amplification
of the noise with signal during the propagation in the fiber.
Noise-noise beat is related to chi-square random variables, which is
independent of the signal, orthogonal to signal-noise beat and
represents the gradual interaction of the Gaussian noise with itself
through the nonlinearity.

\section{Bounds on the Capacity -- Performance of the Half-Gaussian Distribution } \label{sec:fundlimits}

The algebraic model given in the equation (\ref{eq:sim-model}) is a
$2\times2$ MIMO conditionally Gaussian channel model. It is very
closely related to the Rician fading channel model, except that the
term which comes directly from the signal at the output of
(\ref{eq:sim-model}) is the square of the signal, rather than the
signal itself. However it still does not seem amenable to a closed form
expression for the capacity.

For a simple intensity-modulated direct-detection channel
\eqref{eq:imdd}, it is known that half-Gaussian distribution for the
input amplitude
(\emph{i.e.}, a Gaussian distribution
truncated and normalized to nonnegative arguments)
 comes close to the capacity
\cite{mecozzi2001cim,moser2004dbb}.
Motivated by
this, in this section we evaluate the mutual information that the
half-Gaussian distribution $Q^{*}$ can achieve for the zero-dispersion
optical channel, there $Q^{*}$ corresponds to the density function
\[
f_{R_{0}.\Phi_{0}}(r_{0},\phi_{0}) =
\begin{cases} \label{HGUP}
\frac{2}{2\pi\sqrt{2\pi \const{P}}} \exp(-\frac{r_{0}^{2}}{2\const{P}}) & \text{if } r_{0} \geq 0,\\
0 & \text{if } r_{0} < 0.
\end{cases}
\]
We show that, interestingly, a distribution with a truncated Gaussian
profile on the amplitude and uniform phase provides an excellent
global lower bound for the zero-dispersion optical fiber, and is
asymptotically capacity-achieving in a high SNR region where
$\const{P}\gg\max\{\frac{6\pi^2}{\gamma^2\sigma^2\const{L}^3},
\sigma^2\const{L}\}$. Note that the power of $Q^*$ is
$\tilde{\const{P}}=(1-\frac{2}{\pi})\const{P}=0.36\const{P}$.

Although the algebraic model \eqref{eq:sim-model} is much more
tractable than \eqref{eq:zerodis} or the PDF \eqref{eq:pdf},
estimating the resulting output entropy for $Q^*$ is still
complicated, though
$f_{\vecr{Y}_{1},\vecr{Y}_{2}}(\vecd{y}_{1},\vecd{y}_{2})$ can be
computed explicitly. In the case of a simple optical intensity
channel, the data processing inequality for relative entropies was used in
\cite{moser2004dbb} to bound output entropy in terms of input entropy,
by transferring the difficulty to the input side. This technique
however is not immediately implementable for the two-dimensional
problem here. We use several observations based on the algebraic model
to evaluate the mutual information for the distribution $Q^*$.

Let
$f_{\vecr{Y}_{2}|\vecr{X}_{2},\matr{H}}(\vecd{Y}_{2}|\vecd{X}_{2},\matd{H})$
be the conditional probability density function of the phase channel
in the simplified model \eqref{eq:sim-model}. Replacing $\vecr{Z}_2$
with a worst-case Gaussian random variable with the same covariance
matrix for the purpose of the lower bound, we have
\begin{equation*}
 f_{\vecr{Y}_{2}|\vecr{X}_{2},\matr{H}}(\vecd{Y}_{2}|\vecd{X}_{2},\matd{H})=\sum\limits_{k=-\infty}^{\infty}\frac{1}{\sqrt{2\pi\sigma_{\vecr{Y}_{2}}^{2}}}\exp{-
\frac{(\vecd{Y}_{2}+2k\pi-\vecd{X}_{2})^{2}}{2\sigma_{\vecr{Y}_{2}}^{2}}}.
\end{equation*}
where $\sigma_{\vecr{Y}_{2}}^2$ is the variance of $\vecr{Y}_2$, and
the summation is a result of $\textnormal{mod } 2\pi$ reduction.
Pictorially
$f_{\vecr{Y}_{2}|\vecr{X}_{2},\matr{H}}(\vecd{Y}_{2}|\vecd{X}_{2},\matd{H})$
is the summation of shifted Gaussians separated by distance $2\pi$.
However if
$\frac{2}{3}\sigma^2\const{L}\left(\gamma\const{L}\right)^2\const{P}\ll
(2\pi)^{2}$ all Gaussians are localized in the intervals
$[-\frac{(k-1)\pi}{\gamma\const{L}},\frac{(k+1)\pi}{\gamma\const{L}}]$
centered approximately at the mean nonlinear phase noise and only one
Gaussian is present in the interval of interest. In this region we
would be able to find a lower bound by ignoring phase wraparounds.
Conversely, if
$\frac{2}{3}\sigma^2\const{L}\left(\gamma\const{L}\right)^2\const{P}
\gg (2\pi)^{2}$ all Gaussians look globally flat and from the symmetry
of pairwise terms around the middle term, conditional phase tends to
be uniformly distributed in
$[\frac{-(k-1)\pi}{\gamma\const{L}},\frac{(k-1)\pi}{\gamma\const{L}}]$.
In this case, we would be able to lower bound capacity by treating
phase noise as a uniform random variable independent of the channel
input. We make these intuitive statements precise in the next two
subsections.

We also observe that in the zero-dispersion channel, it follows
from phase symmetry that the capacity-achieving input
distribution is
uniform in phase, and thus the search for the optimal
input distribution should
be really done only over one-dimensional distributions. When applying
the input distribution $Q^{*}$ to the original PDF (\ref{eq:pdf}), the
sophisticated dependency on $\phi$ terms disappears since input phase
is assumed to be uniform. We therefore use the exact original PDF
(\ref{eq:pdf}) to find the output entropy, while conditional entropy
is computed from the algebraic model (\ref{eq:sim-model}). Offset term
$\log |\det \matd{J}|$, with $\matd{J}$ being the Jacobian of the
transformation relating these two models, is added to account for the
mismatch between these two models.

\subsection{Capacity bounds in the high-power regime}

In the the high power regime where
$\gamma^{2}\const{P}\sigma^{2}\const{L}^3 \gg 6\pi^{2}$, the
signal-dependent phase noise takes over the entire phase interval
$[0,2\pi]$ and we conclude that phase carries no information. It
follows that in this regime the zero-dispersion model
(\ref{eq:zerodis}) is reduced to the optical intensity-modulated
direct-detection channel \eqref{eq:imdd}. For the latter channel, a
lower bound was derived in \cite{mecozzi2001cim} under average power
constraint which is asymptotically exact. In fact, applying
distribution $Q^*$ to the amplitude PDF \eqref{eq:ampdf}, it is easy
to show that
\begin{equation*}
\const{C}_{H} \geq \frac{1}{2}\log(\rho)-\frac{1}{2},
\end{equation*}
where $\const{C}_H$ is the capacity in the high-power regime, and
$\rho=\const{\tilde{P}}/\sigma^2\const{L}$ is the signal-to-noise
ratio.  Moreover from the duality-based upper bound developed in
\cite{moser2004dbb}
\begin{equation*}
\const{C}_{H} \leq \frac{1}{2}\log(\rho)-\frac{1}{2}+o_{\const{P}}(1),
\end{equation*}
where $o_{\const{P}}(1) \rightarrow 0$ when $\const{P} \rightarrow
\infty$. We conclude that the capacity of the zero-dispersion optical
channel asymptotically, in the region $\const{P}\gg
\max\{\frac{6\pi^2}{\gamma^2\sigma^2\const{L}^3},\sigma^2\const{L}\}$,
is
\begin{IEEEeqnarray*}{rCl}
C\sim\frac{1}{2}\log(\rho)-\frac{1}{2}.
\end{IEEEeqnarray*}
A distribution with a half-Gaussian profile for the amplitude and
uniform phase is capacity achieving at such high powers.

\subsection{Capacity lower bound in the medium-power regime}

As mentioned before, in a power region where
$\gamma^{2}\const{P}\sigma^{2}\const{L}^3 \ll 6\pi^{2}$, the effect of
the phase wraparounds is negligible and the phase channel
qualitatively acts similar to the amplitude channel. The resulting
model is then similar to two amplitude channels correlated with each
other
\begin{IEEEeqnarray}{rCl}\label{eq:sim-mod-g}
\vecr{Y}=\vecr{X}+\matd{H}\vecr{Z}_1+\vecr{Z}_2,
\end{IEEEeqnarray}
where now $\vecr{Y}$ is extended over the entire real line. We proceed to
bound the mutual information for the distribution $Q^*$ and model
\eqref{eq:sim-mod-g}
\begin{IEEEeqnarray*}{rCl}
I(R,\Phi;R_{0},\Phi_{0})|_{Q^{*}}&=&
I(R^{2},\Phi;R_{0}^2,\Phi_{0})|_{Q^{*}}\\
&=&
h(R^{2},\Phi)-h(R^{2},\Phi|R_{0}^{2},\Phi_{0})\\
&=&h(\vecr{Y})-h(\vecr{Y}|\vecr{X}).
\end{IEEEeqnarray*}

It can be shown that $\vecr{Z}_{1} \perp \vecr{Z}_{2}$ and therefore
at the end of the fiber $z=\const{L}$ we can write
\begin{IEEEeqnarray*}{rCl}
\matd{P}&=&\E[(\vecr{Y}-\E\vecr{Y})(\vecr{Y}-\E\vecr{Y})^{H}|\vecr{X}]\nonumber\\&=&
\sigma^{2}\const{L}R_{0}^{2}\left(\begin{array}{ll}
 2 &  1\\
1 & \frac{2}{3}
\end{array}\right)+\sigma^{4}\const{L}^2\left(\begin{array}{ll}
 1 &  \frac{1}{3}\\
\frac{1}{3} & \frac{1}{6}
\end{array}\right),
\end{IEEEeqnarray*}
and
\[
\det
\matd{P}=\frac{\sigma^{4}\const{L}^2}{3}(R_{0}^{4}+R_{0}^{2}\sigma^{2}\const{L}+\frac{1}{6}\sigma^{4}\const{L}^2).
\]

Therefore for the conditional output entropy
\begin{IEEEeqnarray}{rCl}
h(\vecr{Y}|\vecr{X})&=&\log(\gamma \const{L})+\E_{\vecr{X}}[h(\vecr{Y}|\vecr{X}=\vecd{x}_0)] \nonumber \\
&\leq&\log(\gamma \const{L})+\E_{R_{0},\Phi_{0}}[\frac{1}{2}\log((2\pi e)^{2}\det\matd{P})] \label{eq:g-upp-bound} \\
&=&\log(\gamma \const{L})+\log(2\pi
e)+\frac{1}{2}\log(\frac{\sigma^{4}\const{L}^2}{3})\nonumber \\&+&\frac{1}{2}E_{R_{0}}[\log(R_{0}^{2}+\alpha_{1}\sigma^{2}\const{L})(R_{0}^{2}+\alpha_{2}\sigma^{2}\const{L})]\label{eq:cond-output-entropy1},
\end{IEEEeqnarray}
in which $\alpha_{1}=\frac{3+\sqrt{3}}{6}$ and
$\alpha_{2}=\frac{3-\sqrt{3}}{6}$. Note that the entropy of the small
extra non-Gaussian ASE-ASE noise term $\vecr{Z}_2$ was upper bounded
in \eqref{eq:g-upp-bound} by its equivalent worst-case Gaussian. For
the half-Gaussian distribution, with the help of
\cite{gradstejn2000tis} one can verify that
\begin{IEEEeqnarray}{rCl}
\label{eq:amplitude-entropy}
E_{R_{0}}[\frac{1}{2}
\log(R_{0}^{2}+q^{2}\sigma^{2}\const{L})]&=&\frac{1}{2}\log(\const{P})-\frac{1}{2}\log(2)-\frac{1}{2}\zeta
\nonumber \\
&&{-}\:\frac{q^{2}\sigma^{2}\const{L}}{2\const{P}}H([1,1],[\frac{3}{2},2],\frac{q^{2}\sigma^{2}\const{L}}{2\const{P}})\nonumber \\
&&{+}\:\frac{\pi}{2}\textrm{Erfi}(q\sqrt{\frac{\sigma^{2}\const{L}}{2\const{P}}}),
\end{IEEEeqnarray}
where
$\zeta{=}0.5772$
is the Euler constant, $q\in \Reals^+$,
$\textrm{Erfi}(x)=-j\textrm{erf}(jx)$ is the imaginary error function
\cite{gradstejn2000tis}
\begin{IEEEeqnarray*}{rCl}
\textrm{Erfi(x)}= \frac{1}{\sqrt{\pi}}(2x+\frac{2}{3}x^{3}+\frac{1}{5}x^{5}+\frac{1}{27}x^{7}+\ldots), \, |x|\ll1,\IEEEeqnarraynumspace
\end{IEEEeqnarray*}
and $H(\vecd{p},\vecd{q},x)$ is the Hyper-geometric function
\begin{IEEEeqnarray*}{rCl}
H([1,1],[\frac{3}{2},2],x)&=&\sum_{k=0}^{\infty}c_{k}x^{k}=1+\frac{1}{3}x+\frac{4}{45}x^{2}\nonumber \\
&&{+}\:\frac{4}{105}x^{3}+\ldots
\label{eq:hyper-geom} \\
\frac{c_{k+1}}{c_{k}}&=&\frac{k+1}{(k+\frac{3}{2})(k+2)}. \nonumber
\end{IEEEeqnarray*}
From \eqref{eq:cond-output-entropy1}-\eqref{eq:amplitude-entropy}
\begin{IEEEeqnarray}{rCl} \label{cond-entropy}
h(\vecr{Y}|\vecr{X}) &\leq& \log(\const{P}\sigma^{2}\const{L})+\log(\gamma \const{L})+\log(2\pi
e)\nonumber \\ &&-\frac{1}{2}\log(3)-\log(2)-\zeta+F(\frac{\sigma^{2}\const{L}}{2\const{P}},\alpha_{1})\nonumber \\
&&+\:F(\frac{\sigma^{2}\const{L}}{2\const{P}},\alpha_{2}).
\end{IEEEeqnarray}
where
\begin{IEEEeqnarray*}{rCl}
F(x,\alpha)=\frac{\pi}{2}\textnormal{Erfi}(\sqrt{\alpha^2x})-\alpha^2xH([1,1],[\frac{3}{2},2],\alpha^2x).
\end{IEEEeqnarray*}

Output entropy $h(\vecr{Y})$ cannot be straightly computed similar to
$h(\vecr{Y}|\vecr{X})$ from the algebraic model \eqref{eq:sim-mod-g}.
As explained at the beginning of this section, we instead exploit the
phase symmetry of \eqref{eq:pdf} and compute $h(\vecr{Y})$ directly
from the original accurate PDF \eqref{eq:pdf}. The output PDF is computed
as
\begin{IEEEeqnarray}{rCl} \label{frphi}
f_{R,\Phi}(r,\phi) &=&\frac{\exp\left(-\frac{r^{2}}{2\const{P}+\sigma^{2}\const{L}}\right)}{2\pi\sqrt{\pi(2\const{P}+\sigma^{2}\const{L})}}[1+
\textrm{erf}(\sqrt{\frac{2\const{P}}{\sigma^{2}\const{L}(2\const{P}+\sigma^{2}\const{L})}}r)]
\nonumber \\&\approx&
\frac{2}{2\pi}\frac{\exp\left(-\frac{r^{2}}{2\const{P}+\sigma^{2}\const{L}}\right)}{\sqrt{\pi(2\const{P}+\sigma^{2}\const{L})}}
\qquad r \geq 0,
\end{IEEEeqnarray}
where the asymptote of the Bessel function $I_{0}(x) \sim
\frac{1}{\sqrt{2 \pi x}}\exp{x}$ has been used for $|x|\gg 1$, and
$\frac{r}{r_{0}}\approx 1$. Note that this means that with uniform
input phase, from the point of view of input-output densities,
the
zero-dispersion channel acts like the intensity-modulated direct-detection
channel.

From (\ref{frphi}) and changing the variable $\tilde{R}=R^{2}$,
$f_{R^2,\Phi}(r^2,\phi)$ is obtained, which consequently leads to 
\begin{equation} \label{output-entropy}
h(\vecr{Y}){=}\frac{3}{2}\log(\pi)+\log(2\const{P}+\sigma^{2}\const{L})-\frac{1}{2}\zeta+\frac{1}{2}.
\end{equation}

Finally from (\ref{output-entropy}) and (\ref{cond-entropy}), a lower
bound to the capacity of the zero-dispersion channel in the
medium-power range $\sigma^{2}\const{L}{\ll}\const{P}{\ll}
\frac{6\pi^{2}}{\gamma^{2}\sigma^{2}\const{L}^3}$ follows
\begin{IEEEeqnarray}{rCl} \label{lblsnr}
 \const{C}_{M} &\geq& \frac{1}{2}\log(\frac{\const{P}}{\sigma^{2}\const{L}})+\frac{1}{2}\log(\frac{3\pi}{\gamma^{2}
 \const{P}\sigma^{2}\const{L}^3})+\frac{\zeta-1}{2}\nonumber \\
&&{-}\: F(\frac{\sigma^{2}\const{L}}{2\const{P}},\alpha_{1})-F(\frac{\sigma^{2}\const{L}}{2\const{P}},\alpha_{2}),
\end{IEEEeqnarray}
where $\zeta\approx 0.5772$ is the Euler constant.

The first term in (\ref{lblsnr}) is $\frac{1}{2}\log(\rho)$ and is
attributed to the amplitude channel. The rest of the terms are the
phase contributions to the lower bound. Note that the lower bound
(\ref{lblsnr}) depends not only on the ratio
$\rho=\const{P}/\sigma^{2}\const{L}$, but also on the product of the
signal and noise powers. In the low-power regime where $\const{P}\ll
\sigma^{2}\const{L}$ the linear phase cannot be ignored relative to
the nonlinear phase noise.

\section{Two-dimensionality of the capacity}
\label{sec:two-d}

\begin{figure}[tbp]\label{fig:Regions}
\centerline{\includegraphics[scale=0.8333]{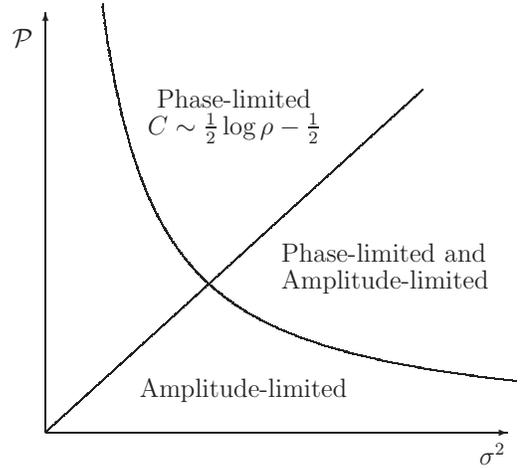}}
\caption{Fundamental transmission regimes in the $\const{P}-\sigma^2$
plane. The line $\const{P}=C_1\sigma^{2}$ marks the boundary between
high and low \SNR\ regimes, where the amplitude channel is
approximately on, or approximately off. The curved boundary is of the
form $\const{P}=\frac{C_2}{\sigma^2}$ and separates two regions where
the phase channel supports, or stops, to support information.}
 \end{figure}

A distinguishing feature of the capacity of the system
(\ref{eq:zerodis}) is the two-dimensionality of the capacity as a
function of both signal and noise powers. Unlike linear Gaussian
channels, the dependency of the capacity of the nonlinear channels to
signal and noise powers is not simply through the ratio
$\rho=\const{P}/\sigma^2\const{L}$. This is in sharp contrast to AWGN
channels where capacity is completely captured by the signal-to-noise
ratio.

For the IM/DD channel \eqref{eq:imdd}, it can be shown that although
channel is still nonlinear in signal and noise, capacity is only a
function of SNR. This basically follows from the scale-invariance
property of the PDF (\ref{eq:ampdf}) with respect to the noise power.
The PDF of the whole channel \eqref{eq:pdf} does not have such
property. By scaling (\ref{eq:zerodis}), if $(Q,V,\gamma)$ is a
solution of (\ref{eq:zerodis}), so is $(\lambda Q,\lambda
V,\frac{\gamma}{\lambda^{2}})$. This implies that changing the Kerr
coefficient, \emph{e.g.}, when $\lambda \rightarrow 0$, keeps the
signal-to-noise ratio constant while capacity varies by $\gamma$. As a
result, capacity is not completely captured by the SNR.
It follows that when characterizing capacity as a function of SNR, one
should make sure that within the working range of parameters, results
are not affected by the choice of $\const{P}$ and $\sigma^{2}$ for a
fixed SNR.

\section{Spectral Considerations} \label{sec:speceff}

Spectral efficiencies of the dispersion-free optical fiber have been
studied in \cite{tang2001scc},\cite{turitsyn2003ico}. In
\cite{tang2001scc}, the author used Pinsker's formula to relate the
spectral efficiency to input-output correlation functions. Such a
result should be used with caution since second-order correlation
functions do not capture the knowledge of the whole PDF, which is
required for the computation of the capacity. In fact, Pinsker's
formula was originally formulated for (linear dispersive) Gaussian
waveform channels. Therefore, the results of \cite{tang2001scc} can be
viewed as a lower bound to the capacity, not ultimate achievable
rates.

The asymptotic tail of the capacity was proved to be growing
unboundedly in \cite{turitsyn2003ico} by an asymptotic analysis. The
authors then concluded that ``a naive straightforward application of
the Pinsker formula for evaluation of the capacity of a nonlinear
channel as, for instance, in \cite{tang2001scc}, can lead to wrong
conclusions regarding the asymptotic behavior of the capacity with
$S/N\rightarrow\infty$''.

There are a number of points to note when comparing the results of
\cite{tang2001scc} with \cite{turitsyn2003ico}. Firstly, while
Pinsker's formula works on waveform channels (like
\eqref{eq:zerodisw}), the finding of \cite{turitsyn2003ico}, is indeed
the \emph{per-sample} capacity of the zero-dispersion channel
\eqref{eq:zerodis}, which is only a lower bound to the capacity of
waveform model \eqref{eq:zerodisw}. Secondly, and most importantly, in
\cite{turitsyn2003ico} the authors neglect the issue of spectrum
broadening, which is essential when comparing capacity in bits per
channel use (as in \cite{turitsyn2003ico}) to spectral efficiency in
bits per second per Hertz (as in \cite{tang2001scc}). Below we discuss
this spectral broadening issue.

The nonlinear term in the phase creates new frequency components in
the pulse spectrum. While the pulse propagates down the fiber, its
spectrum may grow continuously. The amount of spectrum broadening
depends on the pulse shape and generally is proportional to the signal
peak power \cite{agrawal2001nfo}. For the zero-dispersion case,
eventually the pulse may need a large transmission bandwidth when
increasing the average launched power. While bit/symbol versus power
increases indefinitely, bit/sec/Hz may asymptotically vanish with
power, hence having a peak in its curve. In the following, we make an
analogy with FM signals to estimate the bandwidth growth. We assume
that the effect of the noise on bandwidth growth is negligible, and
therefore look at the deterministic spectrum broadening.

The solution of (\ref{eq:zerodis}) in the absence of the noise is
\begin{IEEEeqnarray}{rCl} \label{zdsol} q(t,z)=r_{0}\exp{j(\gamma
r_{0}^{2}z+\phi_{0})}=q(t,0)\exp(j\gamma
|q(t,0)|^{2}z).\IEEEeqnarraynumspace \end{IEEEeqnarray} Since we are
interested in estimates of bandwidth, not the entire spectrum, one can
assume that the input is a single-tone signal whose frequency is the
maximum frequency component of the actual input. The term
$\exp(j\gamma |q(t,0)|^2z)$ individually can be looked upon as an
instance of phase modulation (FM) with no carrier. The spectrum of
$q(t,z)$ in (\ref{zdsol}) then involves Bessel functions and depends
on the envelop of the pulse. 

For pulses of the form 
\begin{IEEEeqnarray*}{rCl}
q(t,0)=e^{j\phi (t)}
\begin{cases}
+A & t\in T_1,\\
-A & t\in T_2,\\
 0 & t\in \bar{T}_1\cap \bar{T}_2,
\end{cases}
\end{IEEEeqnarray*}
where $T_1,\:T_2$ are subsets of $\Reals$ and $\bar{T}$ is the $T$
complement, we have $q(t,\const{L})=e^{j\gamma A^2\const{L}}q(t,0)$.
For these signals, such as constant intensity waveforms, where the
nonlinear phase noise $\phi_{NL}=\gamma r_0^2 z$ is not a function of
time, there is no spectral broadening in the zero-dispersion fiber.
This is a consequence of the fact that the nonlinearity for such
pulses becomes constant across the pulse. For other pulse shapes, a
qualitative argument can be made for the purpose of asymptotic
analysis. Two regimes are considered. In the \emph{narrowband
approximation} regime where the maximum nonlinear phase noise
$\Psi_{\max}=\gamma \const{P}_{0}z$ is less than one radian, spectral
broadening is negligible. If $\const{W}_{z}$ is the bandwidth of the
pulse at distance z, then
\begin{equation*}
\const{W}_{z} \approx 3\const{W}_{0}.
\end{equation*}

In most practical optical systems, $\Psi_{\max}$ exceeds 2$\pi$. In
such a \emph{wideband regime}, the effective single-tone bandwidth,
following Carson's rule, is \begin{equation} \label{carson}
\const{W}_{z} \approx\const{W}_{0}+\Delta f+\const{B}, \end{equation}
in which $\const{B}$ is the bandwidth of
$m(t)=\frac{d}{dt}r_{0}^{2}(t)$, and $\Delta f$ is the frequency
deviation proportional to the peak power of the message $m(t)$. The
precise value of the broadening is pulse-dependent, but from
\eqref{carson} it is qualitatively affine in peak power. Taking the
peak power close to the average power, bandwidth increases linearly
with the average power as well. Because capacity in bit/symbol is at
most logarithmic in power, for such pulse shapes spectral efficiency
should vanish at high average powers.

It follows that the exact  relationship between bits per symbol and
bits per second per Hertz is pulse-dependent. For constant intensity
modulation formats, such as square NRZ, these two are proportional
since there is no bandwidth broadening. All other pulse shapes which
experience even a slight bandwidth enlargement at a given average
power, eventually (when $\const{P}\rightarrow \infty$) require
infinite bandwidth for transmission. For such pulse shapes, the
spectral efficiency asymptotically vanishes as in \cite{tang2001scc}.

The optimal pulse shape from a bits/s/Hz aspect depends on the average
launched power and the target distance. Square pulses, such as RZ or
NRZ pulse formats which are common in optical communication, although
not bandwidth efficient at the transmitter, are optimal at high powers
or long distances. For short distances or lower transmitted power
levels, pulse shapes like Gaussians  which are more spectrally compact
at transmitter generally give better overall spectral efficiency.

As mentioned earlier, with $2W_{0}\const{T}$ complex degrees of
freedom at the input of the fiber, the capacity of the waveform
channel is
$2\const{W}_{0}\const{C}(\frac{\const{P}_0}{2\const{W}_0\const{T}},
\frac{2\sigma_0^{2}\const{W}_\const{L}\const{L}}{2\const{W}_0\const{T}},\gamma)$
bits/sec. If one requires the whole pulse at the end of the fiber to
recover data, then the spectral efficiency of such scheme is
$2\frac{\const{W}_{0}}{\const{W}_{\const{L}}}\const{C}(\frac{\const{P}_0}{2\const{W}_0\const{T}},
\frac{2\sigma_0^{2}\const{W}_\const{L}\const{L}}{2\const{W}_0\const{T}},\gamma)$.
Note that unlike the AWGN channel, bandwidth does not cancel out
through the ratio of the signal and noise powers (\SNR). Spectral
efficiency depends on the initial bandwidth $\const{W}_0$ and
bandwidth enlargement factor
$\frac{\const{W}_{\const{L}}}{\const{W}_{0}}$. For a fixed maximum
input bandwidth, increasing average input power or transmission
distance, not only deteriorates spectral efficiency by a factor of
$\frac{\const{W}_{0}}{\const{W}_{\const{L}}}$, but also allows more
noise in the system, which results in even worse performance.

Note that in our discussion of
spectrum broadening, we have neglected the influence of noise.
In particular, in the presence of the noise even constant intensity
modulation schemes will no longer have a time-independent nonlinear phase,
and hence they will also suffer from spectral broadening.
Such bandwidth enlargement
as a result of the noise might be negligible at low signal and noise
powers, but asymptotically when $\const{P}\rightarrow\infty$ will
cause an infinite spectrum broadening and send the capacity to zero.
In addition to this, in practice a small deviation from an ideal pulse
like NRZ square pulse would lead to the same asymptotic result. 

The spectral efficiencies of the dispersive nonlinear optical fiber
has been studied in \cite{mitra2000nli},
\cite{narimanov2002ccf},\cite{essiambre2010clo},\cite{kahn2004sel}.
It is often reasoned that the peak in the spectral efficiency is the
result of the signal-dependent noise. Increasing the signal power,
amplifies noise power to an extent that sends the capacity to zero. In
contrast, an important result of this paper implies that the peak in
the spectral efficiency, at least at zero-dispersion, is a consequence
of spectrum broadening, not signal-dependent noise. It is a
deterministic product of the nonlinearity, and not a noise property.
From equation (\ref{rse1}) and following Jensen's inequality,
increasing the signal level expands the noise ball as well, but no
more than signal growth rate. 

It is worthwhile to mention that spectrum broadening may be
absent in dispersive fibers. If $\beta_2\neq 0$, the nonlinear
Schr\"odinger equation is integrable and in particular solitons can
exist. These are localized pulses that keep their shape or
periodically recur to their initial state. A fundamental soliton, for
example, suffers from no bandwidth enlargement. It would be
interesting to investigate the relationship between bit/s/Hz and bit/s
in the dispersive fiber. 

Practical application of zero-dispersion optical fibers is quite
limited compared to the standard fiber. Optical fibers can operate
either at shifted zero-dispersion wavelength of 1.55 $\mu$m, or less
commonly, at natural 1.3 $\mu$m zero dispersion wavelength. The
International Telecommunication Union (ITU) Recommendations G.652 and
ITU-T G.653 describe single-mode optical fibers optimized to operate
respectively at 1.31 $\mu$m zero-dispersion and 1.55 $\mu$m shifted
zero-dispersion wavelengths \cite{itu-g652,itu-g653}. There are
nevertheless various issues concerning the practical application of
the nondispersive fibers. Since dispersion is absent, nonlinear
impairments such as self-phase modulation (SPM) and cross-phase
modulation (XPM) might be stronger in nondispersive fibers. This
together with spectrum broadening issue limit the application of
nondispersive fibers in WDM systems. 

Optical fibers greatly benefit from dispersion management. In these
systems fiber segments with positive and negative chromatic dispersion
are placed in tandem to cancel out chromatic dispersion on average.
This keeps pulses localized in their time span, and is known to have
numerous other benefits. The resulting system however might not be
equivalent to the perfectly non-dispersive model discussed in this
paper. In a realistic system, one has a net residual dispersion, loss,
other sources of noise such as Rayleigh scattering
\cite{islam2003rat}, multiple wavelengths and multiple modes which we
have not modeled in this paper.

\section{The structure of the capacity-achieving input distribution}\label{sec:inputdist}

The capacity of the quadrature additive white Gaussian noise channel
(complex AWGN) subject to average and peak power constraints was
proved to be discrete in amplitude with a finite number of mass points
and uniform in phase \cite{shamai1995cap}. Discreteness of the
capacity-achieving input distribution has been established for a
number of other channels, such as the Poisson channel with average and
peak power constraints, the Rayleigh fading channel under average
power constraint \cite{aboufaycal2001cdt} and more generally,
conditionally Gaussian channels under certain conditions
\cite{chan2005cap}. See \cite{chan2005cap} for a list of known
channels having this property.

The amplitude channel in (\ref{rse1}) is closely related to the Rician
fading channel in the wireless communication. The authors of
\cite{aboufaycal2001cdt} proved that the the optimal
capacity-achieving input distribution for the discrete-time memoryless
Rayleigh fading channel is discrete with a finite number of mass
points. The same result was proved in \cite{gursoy2005nrf} for the
non-coherent Rician fading model
\begin{equation*} 
 Y_{i}=mX_{i}+A_{i}X_{i}+Z_{i},
\end{equation*}
where $A_{i}$ and $Z_{i}$ are independent identically-distributed
Gaussian random variables and $m$ is a deterministic constant
representing the line of sight component of the fading.

The IM/DD optical channel shares similarities with the Rician fading
channel. Both have the same type of the signal-dependent noise,
although signal levels are stronger in IM/DD channels. Note that in
fading channels, there is no deterministic rotation or nonlinear phase
noise. The similarity should be understood by the virtue of the
algebraic model (\ref{eq:sim-model}), rather than the original
equation (\ref{eq:zerodis}) or (\ref{zdsol}). Capacity techniques and
coding schemes for Rician fading channels might be useful for IM/DD as
well. In particular, the structure of the capacity-achieving input
distribution tends to be discrete in both cases.

\begin{figure}[tbp]
\centerline{\includegraphics[width=0.95\columnwidth]{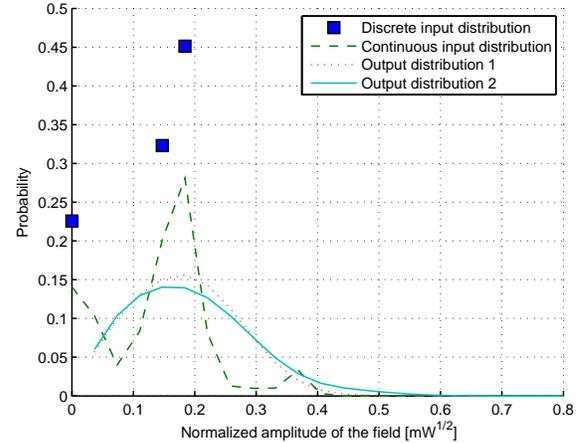}}
\caption{Capacity-achieving input and output distributions at SNR=2.6
dB. The noise power density is 1$\mu$W/GHZ and peak power
$\const{P}_{0}$=10 mW.} \label{fig:DiscInput1}
\end{figure}

A rigorous proof of the discrete character of the capacity-achieving
input distribution for the optical channel is not presented here.
Instead,
we perform a number of simulations to reveal this structure
numerically. Fig.~\ref{fig:DiscInput1}
shows the capacity-achieving input distribution for the amplitude and
the corresponding output distribution. Like the Rician fading channel,
there is always a single mass point at the zero intensity with high
probability. This is not unusual and exists in peak-constrained AWGN
channels as well. Turning off the transmitter sufficiently frequently
helps to stay within the given power budget, while mutual
information is maximized over the remaining degrees of freedom. Like the
AWGN channel, in low \SNR s (SNR $<$6 dB), simple on-off keying (OOK)
is near-optimal. In this case the channel is off more often than it is
on. The ratio of on-to-off probabilities is 0.3 at SNR=2.5 dB, and
decreases with increasing the SNR.

\begin{figure}[tbp]
\centerline{\includegraphics[width=0.95\columnwidth]{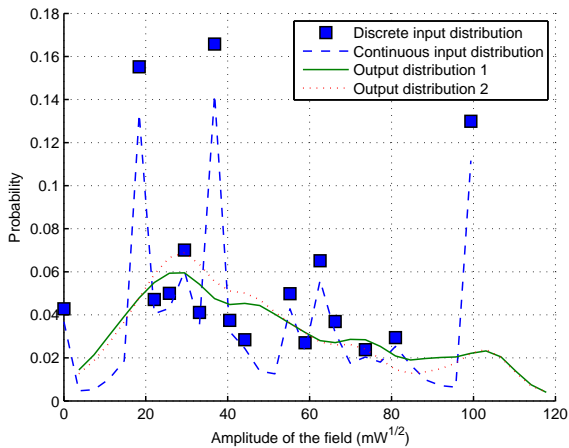}}
\caption{Capacity-achieving
input and output distributions at SNR=13 dB. The noise power density
is 1$\mu$W/GHZ and peak power $\const{P}_{0}$=10 mW}
\label{fig:DiscInput2}
\end{figure}

The surface of the mutual information as a function of the input
probability distribution is flat around the optimum. Hence although
the capacity optimization problem is a convex program with a unique
optimum, there are distributions that come very close to the capacity
but with quite different structure. With increasing SNR, the number
of mass points and their locations increases. More powerful numerical
methods are required to find discrete mass points at higher SNRs. We
use the solution of the interior-point method as an starting point to
search for discrete distributions using a second layer of
interior-point optimization.  The combined method was used to find
discrete mass points at SNR=13 dB in Fig.~\ref{fig:DiscInput2}.
Note that the number of mass points increases considerably
beyond the low power region \SNR $>$6 dB. Some of the points may be
merged with a slight capacity loss. Indeed the controlled increase of
the power shows that new mass points are created by splitting a single
mass point with high probability into two new mass points. With no
peak power, new mass points might come from infinity as conjectured in
\cite{aboufaycal2001cdt} for Rayleigh fading channel under average
power constraint.

It is interesting to note that even though the
capacity-achieving input distribution is not
unique (in the sense that semi-continuous distributions are
near-optimal within our affordable numerical accuracy), the output
distribution appears to be unique. This can be observed in Fig.
\ref{fig:DiscInput1} and Fig.~\ref{fig:DiscInput2}, where the
corresponding output distribution is plotted for a discrete and
semi-continuous input distribution. Capacity deviates $\pm$0.1
bits/symbol which is less than 4\% error.

\section{Conclusion}\label{sec:conclusion}

We have considered the capacity of the per-sample channels
that arise from a model of dispersion-free optical fibers.
The capacity and capacity-achieving input distribution were
evaluated numerically.  We observed that the signal phase carries
little or no information at very low and at very high signal-power
levels, an observation
that
enabled us to find a simple lower bound on the capacity of the
dispersion-free fiber which is asymptotically exact. Although the
overall capacity subject to the average power constraint does not have
a peak in its $\const{C}$-$\const{P}$ curve and grows indefinitely
with input signal power, there exists an optimal power for which the
phase channel reaches its maximum bits/symbol capacity. For
zero-dispersion fibers, neglecting spectral broadening may lead to
wrong conclusions regarding the asymptote of the spectral efficiency.

\appendices \section{A stochastic calculus approach for the derivation
of the conditional PDF} \label{sec:fokkerplanck}

In this appendix we provide a different approach for the derivation of
the conditional PDF of $Q(z)$ in the per-sample channel model
\eqref{eq:zerodis}, namely, $f_{Q(z)|Q(0)}(q(z)|q(0))$. The method is
based on simple mathematical techniques for manipulating random
differential equations, \emph{i.e.}, methods of the stochastic
calculus (see \emph{e.g.}, \cite{gardiner1985hsm}).

\subsection{It\^{o} calculus}
We start by separating the real and imaginary components of $Q(z)$
in \eqref{eq:zerodis}, writing
\begin{IEEEeqnarray}{rCl}\label{eq:x1x2}
  \frac{\partial X_1(z)}{\partial z}&=&-\gamma X_{2}(X_{1}^{2}+X_{2}^{2})+V_{1}(z),  \nonumber\\
  \frac{\partial X_2(z)}{\partial z}&=&\gamma
  X_{1}(X_{1}^{2}+X_{2}^{2})+V_{2}(z),
\end{IEEEeqnarray}
where $X_1(z)=\Re Q(z)$, $X_2(z)=\Im Q(z)$, and $V_i(z)$, $i=1,2$,
are two independent zero-mean real Gaussian processes with
$\E(V_i(z)V_i(z^\prime))=(\sigma^2/2)\delta(z-z^\prime)$.

Note that, in strict mathematical terms, channel model
\eqref{eq:zerodis} (or \eqref{eq:x1x2}) does not exist. To see this,
let $\gamma$ be zero in \eqref{eq:zerodis}. Solution of the equation
is then $W(z)=\int_0^zV(z^\prime)dz^\prime$, \emph{i.e.}, the Wiener
process. The Wiener process is, however, known to be differentiable
nowhere, to satisfy  $\partial W/\partial z=V(z)$. To resolve this
issue, stochastic differential equation \eqref{eq:zerodis} should be
interpreted via its equivalent integral representation
\begin{IEEEeqnarray}{rCl}\label{eq:int-per-sample}
Q(z)=Q_0+\int_0^z \gamma |Q|^2Qdz^\prime+\int_0^z
V(z^\prime)dz^\prime.
\end{IEEEeqnarray}
This eliminates problems with differentiating a stochastic process
(one could also live in the Schwartz space of distributions and
consider the original differential equation in the weak sense).

The system of stochastic differential equations \eqref{eq:x1x2} can be
transformed to polar coordinates via the transformation
\begin{IEEEeqnarray}{rCl} \label{eq:polar}
R=\sqrt{X_1^2+X_2^2},\quad \Phi=\tan^{-1}\frac{X2}{X1}.
\end{IEEEeqnarray}
However since \eqref{eq:x1x2} is a stochastic system, such a
transformation cannot be executed simply based on the ordinary
calculus.

Roughly speaking, the difference between the classical deterministic
calculus and the stochastic calculus stems from the fact that, unlike
the classical calculus in which terms proportional to ${\rm d}t^2$ are
ignored, we can not neglect ${\rm d}W^2$ (square of infinitesimal
increments of the Wiener process) in stochastic equations.
Intuitively, this is because ${\rm
d}W\sim\mathcal{N}_{\Complex}(0,\sigma^2 dz)$ and therefore $\E({\rm
d}W^2)=\sigma^2dz$, \emph{i.e.}, ${\rm d}W^2$ is of order $dz$ and
cannot be neglected. To see this more precisely, let us integrate the
equation of $X_1(z)$ in \eqref{eq:x1x2} from $z$ to $z+\Delta z$ for
small $\Delta z$
\begin{IEEEeqnarray*}{rCl}
  X_1(z+\Delta z)-X_1(z)&\approx& -\gamma X_2(z)\left(X_1(z)^2+X_2(z)^2\right)\Delta z\\&&+\int_{z}^{z+\Delta z} V_1(z^\prime)dz^\prime.
\end{IEEEeqnarray*}
Following the classical calculus, the second term $\int_{z}^{z+\Delta
z}V_1(z^\prime)dz^\prime$ is approximated by $V_1(z)\Delta z$, similar
to the first term. However, it is a fact that the variance of the
summation of a sequence of independent random variables is the
summation of the individual variances. In other words, the linearity
is on the variance, not the standard deviation or the radius of the
balls (signal level). Therefore, the second term properly is
approximated by $V_1(z)\sqrt{\Delta z}$ which consequently affects the
chain rule for differentiation. This can be also understood from the
fact that, when one discretizes the differential equation
\eqref{eq:zerodis} at points $z_k=k\epsilon$, the continuous-space
stochastic process $V(z)$ is replaced by a sequence of random
variables $\left\{\frac{V_k}{\sqrt{\epsilon}}\right\}$,  where $V_k$
is a sequence of i.i.d zero-mean Gaussian random variables with $\E
|V_k|^2=\sigma^2$. 

In \eqref{eq:int-per-sample}, any stochastic integral of the form
$\int_0^zG(Q(z^\prime),z^\prime)V(z^\prime)dz^\prime$ is understood by
definition as
\begin{IEEEeqnarray}{rCl}\label{eq:stieljes-int}
\textnormal{mean-square}\lim\limits_{n\rightarrow\infty}
\sum\limits_{k=1}^nG(Q(\tau_k),\tau_k)\left(W(z^\prime_k)-W(z^\prime_{k-1})\right),
\IEEEeqnarraynumspace
\end{IEEEeqnarray}
where $\textnormal{mean-square}\lim$ is the mean square probabilistic
limit, $z^{\prime}_0=0<z^{\prime}_1<\cdots<z^{\prime}_{n}=z$ is a
partition of the interval $[0,z]$, and $z^\prime_{k-1}\leq \tau_k\leq
z^\prime_k$. Unlike the classical calculus, the choice of the
intermediate point $\tau_k\in[z^\prime_{k-1},z^\prime_k]$ affects the
result of integration. Choosing $\tau_k=z^\prime_{k-1}$ leads to
It\^{o}'s interpretation of the stochastic integral, while using
$\left\{G(Q(\tau_k),\tau_{k-1})+G(Q(\tau_{k-1}),\tau_{k-1})\right\}/2$
in \eqref{eq:stieljes-int} instead of $G(Q(\tau_k),\tau_k)$, gives
Stratonovich's definition. As common in this context, in this paper we
adopt It\^{o}'s definition, which consequently leads to It\^{o}
calculus. 

The following lemma is used when changing variables in a stochastic
differential equation.
\begin{lemma}[It\^{o}'s lemma]\label{lem:ito}
Let $\vecr{x}(z)$ be an n-dimensional stochastic process evolving
according to the following first order stochastic differential
equation
\begin{IEEEeqnarray}{rCl}\label{eq:langevin}
{\rm d}\vecr{X}=\vecd{A}(\vecr{X},z){\rm
d}z+\matd{B}(\vecr{X},z){\rm d}\vecr{W}(z),
\end{IEEEeqnarray}
where $\vecd{A}(\vecr{X},z)\in \Complex^n$ and
$\matd{B}(\vecr{X},z)\in \Complex^{n\times n}$ are respectively
vector-valued and matrix-valued functions, and elements of ${\rm
d}\vecr{W}(z)$ are infinitesimal increments of independent Wiener
processes with
\begin{IEEEeqnarray*}{rCl}
\E\{W_i(z)W_i^*(z^\prime)\}=\min(z,z^\prime),\quad i=1,\cdots,n.
\end{IEEEeqnarray*}
Then for any twice continuously differentiable function
$g(\vecr{x},z)$
\begin{IEEEeqnarray}{rCl}\label{eq:itolemma}
{\rm d}g(\vecr{x},z)&=&\biggl\{\dot{g}(\vecr{X},z)+\sum\limits_i
  \vecd{A}_i(\vecr{X},z)\partial_ig(\vecr{X},z)\nonumber\\
&&+\:\frac{1}{2}\sum\limits_{i,j}[\matd{B}(\vecr{X},z)\matd{B}(\vecr{X},z)^T]_{ij}\partial_i\partial_j
g(\vecr{X},z)\biggr\}{\rm
d}z\nonumber\\&&+\:\sum\limits_{i,j}\matd{B}_{ij}(\vecr{X},z)\partial_ig(\vecr{X},z){\rm
d}W_j(z),
\end{IEEEeqnarray}
where $\dot{g}(\vecr{X},z)=\partial g(\vecr{X},z)/\partial z$, and
$\partial_i$ stands for ordinary partial differentiation with
respect to $\vecr{X}_i$.
\end{lemma}
\begin{IEEEproof}
See \cite {gardiner1985hsm}.
\end{IEEEproof}

In the context of statistical physics, the first order stochastic
dynamical system \eqref{eq:langevin} is called the \emph{nonlinear
Langevin equation}. The quantities $\vecd{a}(\vecr{X},z)$ and
$\matd{B}(\vecr{X},z)$ are \emph{drift} and \emph{diffusion}
coefficients. Note that the term
$\frac{1}{2}\sum\limits_{i,j}[\matd{B}(\vecr{X},z)\matd{B}(\vecr{X},z)^T]_{ij}\partial_i\partial_j
g(\vecr{X},z)$ can be obtained only via the It\^{o} calculus (unless
$\vecd{a}(\vecr{x},z)$ is linear in $\vecr{X}$).

In the case of \eqref{eq:x1x2},
$\matd{B}(\vecr{X},z)=\frac{\sigma}{\sqrt{2}} \matd{I}_2$, and all
other constants are easily extracted from \eqref{eq:x1x2}. Applying
Lemma~\ref{lem:ito} to \eqref{eq:polar}, the per-sample model
\eqref{eq:zerodis} is properly transformed to polar coordinates

\begin{IEEEeqnarray}{rCl}\label{eq:polar-channel}
                 \begin{pmatrix}
                   \frac{\partial R}{\partial z} \\
                   \frac{\partial \Phi}{\partial z} 
                 \end{pmatrix}
               &=&
                 \begin{pmatrix}
                   \overset{\textrm{Ito calculus}}{\overbrace{\frac{\sigma^2}{4R}}} \\
                   \gamma R^2 
                 \end{pmatrix}
               \nonumber\\&&+\:\frac{\sigma}{\sqrt{2}} 
                 \begin{pmatrix}
                   \cos\Phi & \sin\Phi \\
                   -\sin\Phi/R & \cos\Phi/R 
                 \end{pmatrix}
                  \begin{pmatrix}
                   V_{1} \\
                   V_{2} 
                 \end{pmatrix}
               .\IEEEeqnarraynumspace
\end{IEEEeqnarray}
Note that in \eqref{eq:polar-channel} we have assumed 
\begin{IEEEeqnarray}{rCl}
\E\{V_i(z)V_i(z^\prime)\}=\delta(z-z^\prime),\quad i=1,\cdots,n,
\end{IEEEeqnarray}
to be consistent with the statement of the Lemma~\ref{lem:ito}, in
which the amplitude of impulse functions is unit. 

The radial diffusion term $\frac{\sigma^2}{4R}$ is the term which
cannot be obtained by classical calculus and significantly changes the
results. The new channel in the polar coordinates
\eqref{eq:polar-channel} is itself another Langevin equation whose
parameters can be extracted from \eqref{eq:polar-channel}.

\subsection{Fokker-Planck equation}
Consider now the Langevin equation \eqref{eq:polar-channel}, in which
the stochastic process $\vecr{x}(z)=[R,\Phi]^T$ evolves in the
distance dimension $z$. For a fixed $z$, $\vecr{x}(z)$ is a random
variable with a probability density function
$f_{\vecr{x}}(\vecd{x},z)$ parametrized by $z$. Since for the
information-theoretic purposes we are not interested in correlation
between intermediate space samples of random process, \emph{e.g.},
$\E\{\vecr{x}_i(z)\vecr{x}_j(z^\prime)^{H}\}$,
$0<z<z^\prime<\const{L}$, a single conditional PDF at the output of
the fiber completely describes the underlying channel. The stochastic
process contains much more information, but this is irrelevant to our
application.

Let $g(\vecr{x},z)$ be a general function in Lemma~\ref{lem:ito},
independent of $z$, and with vanishing boundary terms in $\vecr{x}_i$
(\emph{i.e.}, if the support of densities extends to infinity, then
$g(\pm\infty,z)=0$). One can look at the evolution of $\E
g(\vecr{X},z)$, which is deterministic. The result is a deterministic
equation in terms of $g(\vecr{x},z)$ and $f_{\vecr{x}}(\vecd{x},z)$.
Since $g$ can be varied to be any change of variable, it follows that
$f_{\vecr{x}}(\vecd{x},z)$ should satisfy a certain evolution
equation. Let, therefore, fix $\vecr{X}=\vecd{x}$ in
\eqref{eq:itolemma}, multiply both sides of \eqref{eq:itolemma} by
$f_{\vecr{x}}(\vecd{x},z)$, \emph{i.e.}, the PDF of $\vecr{x}(z)$ at a
fixed $z$, and integrate with respect to $\vecd{x}$. Both sides can
then be integrated by parts which transfers differentials from
$g(\vecd{x},z)$ to $f_{\vecr{x}}(\vecd{x},z)$. Since the resulting
integral holds for any $g(\vecd{x},z)$, the following lemma is
obtained (for the case $n=1$).

\begin{lemma}[Nonlinear Fokker-Planck equation]
Let $f_{\vecr{x}}(\vecd{x},z)$ be the probability density function of
$\vecr{x}(z)$ at a fixed $z$ in the Langevin equation
\eqref{eq:langevin}. Then $f_{\vecr{x}}(\vecd{x},z)$ satisfies the
following differential equation
\begin{IEEEeqnarray}{rCl}\label{eq:fp-g}
\frac{\partial f_{\vecr{X}}(\vecd{x},z)}{\partial
  z}&=&-\sum\limits_i\frac{\partial}{\vecd{x}_i}[\vecd{a}(\vecd{x},z)f_{\vecr{x}}(\vecd{x},z)]\nonumber\\
&&+\:\frac{1}{2}\sum\limits_{i,j}\frac{\partial^2}{\partial\vecd{x}_i\partial\vecd{x}_j}\left
 \{[\matd{B}(\vecd{X},z)\matd{B}(\vecd{X},z)^T]_{ij} f_{\vecr{x}}(\vecd{x},z)\right\}.\IEEEeqnarraynumspace
\end{IEEEeqnarray}
\end{lemma}
\begin{IEEEproof}
For the single variable case, the proof outlined above simply gives
the desired result. The generalization to the multivariable case is,
however, complicated since boundary terms are now surfaces and curves,
instead of points. See \cite{gardiner1985hsm} for the complete proof.
\end{IEEEproof}

Description of the probability density function via the Fokker Planck
equation is incomplete without specifying the boundary conditions. In
the multivariable case, such conditions may render the problem hard to
solve analytically. In this paper, we assume
\begin{IEEEeqnarray*}{rCl}
f_{\vecr{X}}(\vecd{x},z=0)=\delta(q-q_0),\quad
\lim\limits_{|\vecd{x}|_i\rightarrow
\infty}f_{\vecr{X}}(\vecd{x},z)=0,\quad \forall i.
\end{IEEEeqnarray*}
The delta function implies that we are looking for the conditional PDF
$f_{\vecr{X}}(\vecd{x},z|\vecd{x}_0,z_0)$. In the following, we
unambiguously use $f_{\vecr{X}}(\vecd{x},z)$ instead of
$f_{\vecr{X}}(\vecd{x},z|\vecd{x}_0,z_0)$ for the sake of brevity.
Depending on the structure of the problem, other appropriate boundary
conditions might be assumed, such as periodic, absorbing or reflecting
boundary conditions.

Note that according to the Fokker-Planck equation, matrix
$\matd{B}(\vecr{X},z)$ in the dynamics of the stochastic process
affects the PDF in the form of
$\matd{B}(\vecr{X},z)\matd{B}(\vecr{X},z)^T$. In other words,
probabilistically channel \eqref{eq:polar-channel} is equivalent to
\begin{IEEEeqnarray}{rCl}\label{eq:polar-channel-sim}
                 \begin{pmatrix}
                   \frac{\partial R}{\partial z} \\
                   \frac{\partial \Phi}{\partial z} 
                 \end{pmatrix}
               &=&
                 \begin{pmatrix}
                   \frac{\sigma^2}{4R} \\
                   \gamma R^2 \\
                 \end{pmatrix}
               +\:\frac{\sigma}{\sqrt{2}} 
                 \begin{pmatrix}
                   1 & 0 \\
                   0 & \frac{1}{R^2} 
                 \end{pmatrix}
                 \begin{pmatrix}
                   V_{1} \\
                   V_{2} 
                 \end{pmatrix}, \end{IEEEeqnarray}
which corresponds to the following Fokker-Planck equation
\begin{IEEEeqnarray}{rCl}\label{eq:fp}
\frac{\partial f(r,\phi,z)}{\partial z}&=&-\gamma
r^{2}\frac{\partial f}{\partial
\phi}+\frac{1}{4}\sigma^2\left[\frac{\partial^{2}f}{\partial
r^{2}}+\frac{1}{r^{2}}\frac{\partial^{2}f}{\partial \phi^{2}}
\right]\nonumber\\
&&-\:\overset{\textrm{Ito term}}{\overbrace{\frac{\sigma^2}{4}\frac{\partial}{\partial r}\left(\frac{f}{r}\right)}},\IEEEeqnarraynumspace\\
f(r,\phi,0)&=&\delta(r-r_0,\phi-\phi_0). \nonumber
\end{IEEEeqnarray}
\begin{remark}[Phase Symmetry]
If $f(r,\phi,z)$ is a solution of \eqref{eq:fp}, so is
$f(r,\phi-\phi_1,z)$ for any $\phi_1$. The PDF is therefore symmetric
with respect to the phase, \emph{i.e.}, it is a function of
$\phi-\phi_0$. This essentially comes from the fact that in the
nonlinear Schr{\"o}dinger equation, the cubic nonlinearity is of the
form $|Q|^2Q$, as opposed to $Q^3$. This roots back to the underlying
physics of the fiber, in which the nonlinear refractive response of
the silica glass to an external light beam is deterministically
proportional to the intensity $|Q|^2$ of the incoming beam. The
absolute value is responsible for some very important properties of
the nonlinear Schr{\"o}dinger equation.
\end{remark}

We proceed to solve the resulting partial differential equation
\eqref{eq:fp}. First, we find the marginal PDF of the amplitude
channel alone. Integrating both sides of \eqref{eq:fp} with respect to
$\phi$ in the interval $[0,2\pi]$, assuming phase continuity
$f(r,0,z)=f(r,2\pi,z)$, and performing phase marginalization, we
obtain
\begin{IEEEeqnarray}{rCl}\label{eq:amp-pde}
\frac{\partial f(r,z)}{\partial
z}&=&\frac{1}{4}\sigma^2\frac{\partial^{2}f(r,z)}{\partial r^{2}}
-\frac{\sigma^2}{4}\frac{\partial}{\partial r}\left[\frac{f(r,z)}{r}\right],\IEEEeqnarraynumspace\\
f(r,0)&=&\delta(r-r_0).\nonumber
\end{IEEEeqnarray}
The PDF of the amplitude, $f(r,z)$, satisfies \eqref{eq:amp-pde}. If
at $z=0$, $r=r_0$, then the conditional PDF $f(r,z|r_0)$ is the
solution of \eqref{eq:amp-pde}. We make the following observations
about the resulting PDE.

\begin{remark}[Scaling Property]\label{rem:amp-pdf-scaling}
Partial differential equation \eqref{eq:amp-pde} admits an important
scaling property: if $f(r,z|r_0)$ is a solution of \eqref{eq:amp-pde},
then so is $f(\lambda r,\lambda^2z|\lambda r_0)$ for any real nonzero
$\lambda$. Such scaling is indeed a \emph{Lie symmetry group} of the
partial differential equation \eqref{eq:amp-pde} and corresponds to a
\emph{conserved quantity}.  The essential feature of the symmetry
group is that it conserves the set of solutions of the differential
equation, inducing a set of orbits under action by the group. Each
symmetry group may be visualized as permuting integral curves of the
partial differential equation among themselves.
\end{remark}
\begin{remark}[Robustness]\label{rem:robustness}
The PDF of the amplitude does not depend on the Kerr nonlinearity
constant $\gamma$, hence the operation of an amplitude detector
(\emph{e.g.}, a photodetector) is also independent of $\gamma$.
\end{remark}

We use the scaling  property in Remark~\ref{rem:amp-pdf-scaling} to
solve the partial differential equation \eqref{eq:amp-pde}. Such
scaling implies that the ratios $r^2/z$ and $rr_0/z$ are important
quantities for the equation \eqref{eq:amp-pde}. We therefore search
for the solutions of the form
$f(r,z)=h(\frac{r^2}{z},\frac{rr_0}{z})$. Substituting this into
\eqref{eq:amp-pde} and after straightforward algebra, we obtain the
solution
\begin{IEEEeqnarray}{rCl}\label{eq:amplitude-pdf}
f(r,z)=\frac{2r}{\sigma^2z}e^{-\frac{r^2+r_0^2}{\sigma^2
z}}I_0\left(\frac{rr_0}{\sigma^2z}\right),
\end{IEEEeqnarray}
where we have scaled $f$ so that it stands for a probability
distribution, and $f\rightarrow \delta(r-r_0)$ when $z \rightarrow 0$
( if $z \rightarrow 0$, then $I_0(\frac{2rr_0}{\sigma^2 z})\rightarrow
\frac{1}{r}\sqrt{\frac{\sigma^2 z}{2\pi}}e^{\frac{2rr_0}{\sigma^2
z}}$). Below we use \eqref{eq:amplitude-pdf} to solve \eqref{eq:fp}.
Note that alternatively one may solve \eqref{eq:amp-pde} using Remark
\ref{rem:robustness}: since the PDF of the amplitude does not depend
on $\gamma$, one can assume $\gamma=0$ in the original equation
$\eqref{eq:zerodis}$ and find the PDF of the amplitude of a complex
Brownian motion.

We use the method of \emph{separation of variables} to solve the PDE
\eqref{eq:fp}. Assuming $f$ is separable in $r$ and  $\phi$,
\emph{i.e.},
\begin{IEEEeqnarray}{rCl}\label{eq:f=gh}
f(r,\phi,z)=g(r,z)h(\phi,z),
\end{IEEEeqnarray}
and plugging \eqref{eq:f=gh} into \eqref{eq:fp}, we get
 \begin{IEEEeqnarray}{rCl}\label{eq:sepvar}
&&\frac{1}{g}\frac{\partial g(r,z)}{\partial
z}+\frac{1}{h}\frac{\partial h(\phi,z)}{\partial
  z}=-\gamma r^2\frac{1}{h(\phi,z)}\frac{\partial h(\phi,z)}{\partial
  \phi}\nonumber\\
&&+\:\frac{1}{4}D\biggl\{\frac{1}{g(r,z)}\frac{\partial^{2}g(r,z)}{\partial
r^{2}}+\frac{1}{r^{2}}\frac{1}{h(r,z)}\frac{\partial^{2}h(\phi,z)}{\partial \phi^{2}}\nonumber\\
&& -\frac{1}{g(r,z)}\frac{\partial }{\partial
r}(\frac{g(r,z)}{r})\biggr\}.
\end{IEEEeqnarray}
If
\begin{IEEEeqnarray}{rCl}\label{eq:hc}
\frac{1}{h(\phi,z)}\frac{\partial h(\phi,z)}{\partial
  \phi}=c
\end{IEEEeqnarray}
for some constant $c$, then
\begin{IEEEeqnarray*}{rCl}
\frac{1}{h(\phi,z)}\frac{\partial^2 h(\phi,z)}{\partial
  \phi^2}=c^2.
\end{IEEEeqnarray*}
In this case, \eqref{eq:sepvar} allows a separation of variables.
Solution of the phase part in \eqref{eq:hc} is $h(\phi,z)=\exp
\left[c\left(\phi-\phi_0\right)\right]$ and the equation for the
amplitude part, $g(r,z)$, is
\begin{IEEEeqnarray}{rCl}\label{eq:gc}
\frac{\partial g_c}{\partial z}=\frac{\sigma^2}{4}\frac{\partial^2
g_c}{\partial
  r^2}+\left(\frac{c^2}{r^2}\frac{\sigma^2}{4}-\gamma
  cr^2\right)g_c-\frac{\sigma^2}{4}\frac{\partial}{\partial r}\left(\frac{g_c}{r}\right).\IEEEeqnarraynumspace
\end{IEEEeqnarray}
Due to phase periodicity, naturally one can assume $c = jm$, whence the
PDF is written as a Fourier series with coefficient $g_m(z,r)$.

Although \eqref{eq:gc} is linear in $g_c$, applying Fourier or Laplace
transform gives another PDE which is as hard as \eqref{eq:gc} to
solve. We prefer to use a \emph{variational method}. If $c=0$,
\eqref{eq:gc} is reduced to \eqref{eq:amp-pde} whose solution is then
\eqref{eq:amplitude-pdf}. We therefore assume the general solution for
$c\neq 0$ to be a variation of the solution at $c=0$,
\begin{IEEEeqnarray}{rCl}\label{eq:gc-candid}
g_c(r,z)=\frac{r}{\pi
\sigma^2z}e^{-\left(r^2+r_0^2\right)a(z)}I_m(2rr_0b(z))
\end{IEEEeqnarray}
for some unknown $a(z)$ and $b(z)$ to be determined. Finally,
$\eqref{eq:gc-candid}$ is plugged in \eqref{eq:gc} to get differential equations for $a(z)$ and $b(z)$. After some rather
tedious but straightforward algebra, the overall PDF
\begin{IEEEeqnarray*}{rCl}
f(r,z)=\sum\limits_{m=-\infty}^{\infty}e^{jm(\phi-\phi_0-\gamma r_0^2z)}g_m(r,z)
\end{IEEEeqnarray*}
is obtained, in which
\[
g_m(r,z)=\frac{rb(m)}{\pi}e^{-a_m(z)\left(r^2+r_0^2\right)}I_{|m|}(2b_m(z)r_0r),
\]
where
\begin{IEEEeqnarray*}{rCl}
a_m(z)&=&\frac{\sqrt{jm\gamma}}{\sigma}\coth\sqrt{j m\gamma \sigma^2 }z,\nonumber \\
b_m(z)&=&\frac{\sqrt{jm\gamma}}{\sigma}\frac{1}{\sinh\sqrt{jm\gamma\sigma^2}z}.
\end{IEEEeqnarray*}

Functions $g_m(r,z)$ can be considered as eigenfunctions in the
amplitude PDE \eqref{eq:gc}, $\partial g/\partial z=Hg=\lambda g$ (see
the literature in time-dependent Schr\"odinger equation in quantum
mechanics). The PDF \eqref{eq:pdf} then appears to be an expansion in
terms of eigenfunctions of the associated Fokker-Planck equation. Each
of these terms is a solution of the Fokker-Planck equation and the
summation in \eqref{eq:pdf} is to create the delta function
$\delta(\phi-\phi_0)$ at the beginning of the fiber.

As this paper was going to press, we became aware that a similar
Fokker-Planck approach to describing the statistics of the model
\eqref{eq:zerodis} has been worked out---in the
context of nonlinear oscillators---in \cite{prilepsky2005snd}.

\section{Useful identities}
\label{sec:identities}
\begin{IEEEeqnarray}{rCl}
\label{eq:Im}
\frac{1}{2\pi}\int\limits_{0}^{2\pi}e^{-jm\theta+x\cos(\theta-\theta_0)}d\theta &=&I_m(x)e^{-jm\theta_0}
\\
\label{eq:Im1-Im2}
\int\limits_{0}^{\infty}xe^{-ax^2}I_m(bx)I_m(cx)dx&=&\frac{1}{2a}e^{\frac{b^2+c^2}{4a}}I_m(\frac{bc}{2a})
\end{IEEEeqnarray}


\begin{IEEEbiographynophoto}{Frank R. Kschischang} received the B.A.Sc. degree (with honors) from
the University of British Columbia, Vancouver, BC, Canada, in 1985 and
the M.A.Sc.  and Ph.D. degrees from the University of Toronto,
Toronto, ON, Canada, in 1988 and 1991, respectively, all in electrical
engineering.  He is a Professor of Electrical and Computer Engineering
at the University of Toronto, where he has been a faculty member since
1991. During 1997-98, he was a visiting scientist at MIT, Cambridge,
MA and in 2005 he was a visiting professor at the ETH, Zurich.

His research interests are focused primarily on the area of channel
coding techniques, applied to wireline, wireless and optical
communication systems and networks.  In 1999 he was a recipient of the
Ontario Premier's Excellence Research Award and in 2001 (renewed in
2008) he was awarded the Tier I Canada Research Chair in Communication
Algorithms at the University of Toronto.  In 2010 he was awarded the
Killam Research Fellowship by the Canada Council for the Arts.
Jointly with Ralf Koetter he received the 2010 Communications Society
and Information Theory Society Joint Paper Award.  He is a Fellow of
IEEE and of the Engineering Institute of Canada.
  
During 1997-2000, he served as an Associate Editor for Coding Theory
for the IEEE TRANSACTIONS ON INFORMATION THEORY. He also served as
technical program co-chair for the 2004 IEEE International Symposium
on Information Theory (ISIT), Chicago, and as general co-chair for
ISIT 2008, Toronto.  He served as the 2010 President of the IEEE
Information Theory Society.
\end{IEEEbiographynophoto}

\end{document}